\documentclass[aps,pra,twocolumn,showpacs]{revtex4}
\usepackage{bbm,amsmath,amssymb,amsbsy,graphicx,subfigure,times}
\usepackage[latin1]{inputenc}
\usepackage{cerchio}
\usepackage{color}

\newcommand{\arcsinh}{{\rm arcsinh}}

\newcommand{\R}{\mathbbm{R}}

\newcommand{\id}{\mathbbm{1}}

\renewcommand{\det}{{\rm Det}\,}
\newcommand{\gr}[1]{\boldsymbol{#1}}
\newcommand{\be}{\begin{equation}}
\newcommand{\ee}{\end{equation}}
\newcommand{\bea}{\begin{eqnarray}}
\newcommand{\eea}{\end{eqnarray}}
\newcommand{\ket}[1]{|#1\rangle}
\newcommand{\bra}[1]{\langle#1|}
\newcommand{\ketbra}[2]{\vert #1 \rangle \! \langle #2 \vert}

\newcommand{\N}{{\cal N}}

\newcommand{\sig}{\gr{\sigma}}
\newcommand{\eps}{\gr{\varepsilon}}

\newcommand{\abs}[1]{\left\vert#1\right\vert}
\newcommand{\eq}[1]{Eq.~(\ref{#1})}
\newcommand{\ineq}[1]{Ineq.~(\ref{#1})}
\newcommand{\eg}{\emph{e.g.}~}
\newcommand{\ie}{\emph{i.e.}~}

\begin{document}

\title{Multipartite entanglement in three-mode Gaussian states of continuous \\ variable
systems: Quantification, sharing structure, and decoherence}

\author{Gerardo Adesso$^{1,2}$, Alessio Serafini$^{3,4}$ and
Fabrizio Illuminati$^1$}

 \affiliation{$^1$Dipartimento di Fisica ``E. R.
Caianiello'', Universit\`a degli Studi di Salerno; CNR-Coherentia,
Gruppo di Salerno; and INFN Sezione di Napoli-Gruppo Collegato di
Salerno; Via
S. Allende, 84081 Baronissi (SA), Italy \\
$^2$Centre for Quantum Computation, DAMTP, Centre for Mathematical
Sciences, University of Cambridge, Wilberforce Road, Cambridge CB3
0WA, United Kingdom \\
$^3$Institute for Mathematical Sciences, Imperial College London,
SW7 2PE, United Kingdom; \\ and QOLS, The Blackett Laboratory,
Imperial
College London, Prince Consort Road, SW7 2BW, United Kingdom \\
$^4$Department of Physics \& Astronomy, University College London,
Gower Street, London WC1E 6BT, United Kingdom}

\pacs{03.67.Mn, 03.65.Ud}

\begin{abstract}
We present a complete analysis of multipartite entanglement of
three-mode Gaussian states of continuous variable systems.
We derive standard forms which characterize the covariance matrix of
pure and mixed three-mode Gaussian states up to local unitary
operations, showing that the local entropies of pure Gaussian states
are bound to fulfill a relationship which is stricter than the
general Araki-Lieb inequality. Quantum correlations can be
quantified by a proper convex roof extension of the squared
logarithmic negativity, the continuous-variable tangle, or {\em
contangle}. We review and elucidate in detail the proof that in
multimode Gaussian states the contangle satisfies a monogamy
inequality constraint [G. Adesso and F. Illuminati, New J. Phys. 8,
{\bf 15} (2006)]. The residual contangle, emerging from the monogamy
inequality, is an entanglement monotone under Gaussian local
operations and classical communication and defines a measure of
genuine tripartite entanglement. We determine the analytical
expression of the residual contangle for arbitrary pure three-mode
Gaussian states and study in detail the distribution of quantum
correlations in such states. This analysis yields that pure,
symmetric states allow for a promiscuous entanglement sharing,
having both maximum tripartite entanglement and maximum couplewise
entanglement between any pair of modes. We thus name these states
GHZ/$W$ states of continuous variable systems because they are
simultaneous continuous-variable counterparts of both the GHZ and
the $W$ states of three qubits. We finally consider the effect of
decoherence on three-mode Gaussian states, studying the decay of the
residual contangle. The GHZ/$W$ states are shown to be maximally
robust against losses and thermal noise.
\end{abstract}

\date{March 30, 2006}

\maketitle

\section{Introduction}

Multipartite entanglement is one of the most fundamental
and puzzling aspects of quantum mechanics and quantum
information theory. Although some progress has been recently
gained in the understanding of the subject, many basic problems
are left to investigate in this fascinating area of research.
Multipartite entanglement poses a basic challenge both for the
obvious reason that it is ubiquitous to any practical realization
of quantum communication protocols and quantum computation
algorithms, and because of its inherent, far-reaching
fundamental interest \cite{chuaniels,book1}.

The steps undertaken so far in the attempt to reach some
understanding of quantum entanglement in multipartite settings can
be roughly classified in two categories. On the one hand, the
{\em qualitative} characterization of multipartite entanglement can
be investigated exploring the possibility of transforming a
multipartite state into another under different classes of local
transformations and introducing distinct equivalence classes of
multipartite entangled states \cite{book1}. On the other hand, a {\em
quantitative} characterization of the entanglement of states shared
by many parties can be attempted: this approach has lead to the
discovery of so-called {\it monogamy inequalities}, constraining the
maximal entanglement shared by different internal partitions of a
multipartite state \cite{CKW,Osborne}.
Such inequalities are uprising as one of the possible
fundamental guidelines on which proper measures of
multipartite entanglement should be built.

Recently, much effort has been devoted to the study of
entanglement in continuous-variable systems, focusing
both on quantum communication protocols and on fundamental
theoretical issues \cite{CVbook1, CVbook2, review, eisplenio}.
A rich and complex structure has emerged,
already in the restricted, but physically relevant,
context of Gaussian states. The generic study of Gaussian states
presents many interesting and appealing features, because it can
be carried out exploiting the powerful formalism based on covariance
matrices and symplectic analysis. These properties allow to face and answer
questions that are in general much harder to discuss in discrete variable
systems, and open up the possibility to shed some light upon general
facets of multipartite entanglement, that might carry over to systems
of qubits and qudits.

For two-mode Gaussian states, the qualification and quantification
of bipartite entanglement have been intensively
studied, and a rather complete and coherent understanding
begins to emerge \cite{prl,extremal}.
However, in the case of three-mode Gaussian states, the simplest
non-trivial instance of multi-party entangled Gaussian states
that can be conceived, the multipartite sharing structure
of quantum correlations presents several subtle structural
aspects that need to be elucidated.
Therefore, three-mode Gaussian states constitute an elementary
but very useful theoretical laboratory that is needed toward the
understanding of the patterns by which quantum correlations
distribute themselves among many parties.

A fairly complete qualitative characterization of entanglement in
three-mode Gaussian states has been recently achieved \cite{kraus}.
In the present paper, we study and present a fully quantitative
characterization of entanglement in three-mode Gaussian states. We
discuss the general properties of bipartite entanglement in pure and
mixed states as well as the definition and determination of monogamy
inequalities, genuine tripartite entanglement, and the ensuing
structure of entanglement sharing. We single out a special class of
pure, symmetric, three-mode Gaussian states that are the
continuous-variable analogues and possess the same entanglement
properties of both the $W$ and the Greenberger-Horne-Zeilinger (GHZ)
maximally entangled states of three qubits. Finally, we discuss the
decoherence of three-mode Gaussian states and the decay of
tripartite entanglement in the presence of noisy environments, and
outline different possible generalizations of our results to
$n$-mode Gaussian states with arbitrary $n$.

The paper is organized as follows. In Section II we provide
a self-contained introduction to the symplectic formalism
for covariance matrices, and review the structure of
entanglement in two-mode Gaussian states. In Section III
we apply the known facts on two-mode states and the symplectic
formalism to provide a systematic quantification of bipartite
entanglement in three-mode Gaussian states. In Section IV
we review the concept of continuous-variable tangle and
the continuous-variable monogamy inequalities recently
derived \cite{contangle,sharing}, and exploit these results to
quantify the genuine tripartite entanglement in three-mode
Gaussian states. In Section V we analyze the distributed
entanglement and the structure of entanglement sharing in
three-mode Gaussian states, and identify some classes
of symmetric, pure and mixed, three-mode Gaussian states with
special entanglement properties, including the so-called
``GHZ/$W$'' states that maximize simultaneously the genuine
tripartite entanglement and the bipartite entanglement of any
two-mode reduction. In Section VI we discuss the decoherence
of three-mode Gaussian states and the decay of tripartite
entanglement due to the coupling with the environment.
Finally, in Section VII we give some concluding remarks and
sketch an outlook on some future developments and extensions
to more general states and instances of continuous-variable
systems.


\section{Preliminary facts and definitions for Gaussian states}

In this section, we will introduce basic facts and notation about
Gaussian states of bosonic fields, reviewing some of the existing
separability criteria for two-mode and multimode states and the
computable measures of entanglement available for bipartite systems.
Such basic results will be needed in extending the analysis to
multipartite quantum correlations in multimode Gaussian states.

\subsection{Covariance matrices, symplectic eigenvalues, and inseparability criteria}

Let us consider a quantum system described by $n$ pairs of
canonically conjugated operators, for instance the quadrature
operators of a bosonic field, $\{\hat x_j,\hat p_j\}$, satisfying
the canonical commutation relations $[\hat x_{j},\hat
p_{k}]=\delta_{jk}$. For ease of notation, let us define the vector
of field operators $\hat R = \{\hat x_1,\hat p_1,\ldots,\hat
x_n,\hat p_n\}$ and note that the commutation relations can be
written as $[\hat R,\hat R]=2i\Omega$, where the symplectic form
$\Omega$ is defined as \be \Omega=\bigoplus_{1}^{n}
\gr\omega\,,\quad \gr\omega=\left(\begin{array}{cc}
0&1\\
-1&0
\end{array}\right) \; , \label{symform}
\ee where $\bigoplus$ denotes the direct sum. Any state of such a
system is represented by a hermitian, positive, trace-class operator
$\varrho$, the so-called density matrix. Gaussian states are defined
as states with Gaussian characteristic (and quasi-probability)
functions: a state $\varrho$ is Gaussian if and only if its
characteristic function \be\chi(\xi) \equiv {\rm Tr}\,[\varrho
D_{\xi}]\,,\ee where $\xi\in{\mathbbm R}^{2n}$ is a real vector and
$D_{\xi}=\exp{(i\hat R^{\sf T}\Omega \xi)}$ is Glauber's
displacement operator, is a multivariate Gaussian in the variable
$\xi$. This definition implies that a Gaussian state $\varrho$ is
completely determined by the vector $X$ of its first moments of the
field operators, whose entries are given by $X_j=\,{\rm Tr}[\varrho
\hat R_j]$, and by the covariance matrix (CM) $\gr\sigma$, whose
entries $\sigma_{jk}$ are given by \be \sigma_{jk} = \,{\rm
Tr}\,[\varrho (\hat R_{j}\hat R_k+\hat R_{k}\hat R_j)]/2 - X_j X_k
\; . \ee Explicitly, the characteristic function $\chi(\xi)$ of a
Gaussian state with first moments $X$ and CM $\sig$ is given by \be
\chi(\xi) = \,{\rm e}^{-\frac12 \xi^{\sf T}\Omega^{\sf
T}\sig\Omega\xi+i X^{\sf T}\Omega\xi} \, . \label{gauss}\ee Gaussian
states play a prominent role in practical realizations of
continuous-variable (CV) quantum information protocols. They can be
created and manipulated with relative ease with current technology
\cite{francamentemeneinfischio}, and, thanks to their simple
description in terms of covariance matrices, provide a powerful and
relevant theoretical framework for the investigation of fundamental
issues.

All the unitary operations mapping Gaussian states into Gaussian
states are generated by polynomials of the first and second order in
the quadrature operators. First order operations are just
displacement operators $D_{\xi}$, which leave the CM unchanged while
shifting the first moments. Such unitary operations, by which first
moments can be arbitrarily adjusted, are manifestely local: this
entails that first moments can play no role in the entanglement
characterization of CV states and will be thus henceforth neglected,
reducing the description of the states under exam to the CM
$\gr\sigma$. On the other hand, unitary operations of the second
order act, in Heisenberg picture, linearly on the vector $\hat R$:
$\hat R \mapsto S\hat R$, where the matrix $S$ satisfies $S^{\sf
T}\Omega S=\Omega$. The set of such (real) matrices form the {\em
real symplectic group} $Sp_{2n,\R}$ \cite{folland,pramana}.
Therefore, these unitary operations are called symplectic
operations. Symplectic operations act on a CM $\gr\sigma$ by
congruence: $\gr\sigma\mapsto S^{\sf T} \gr\sigma S$.

Besides describing most unitary Gaussian operations currently
feasible in the experimental practice (namely beam-splitters,
squeezers, and phase-shifters), the symplectic framework is
fundamental in the theoretical analysis of CMs:
for any physical CM $\gr\sigma$ there exist a symplectic
transformation $S\in
Sp_{2n,\R}$ such that $S^{\sf T}\gr\sigma S=\gr\nu$, where
$$
\nu=\oplus_{j=1}^{n} \,{\rm diag}\, (\nu_j,\nu_j) \; .
$$
The quantities $\{\nu_j\}$, uniquely determined for every CM
$\gr\sigma$, are referred to as the {\it symplectic eigenvalues} of
$\gr\sigma$, while $\gr\nu$ is said to be the {\it Williamson normal
form} associated to $\gr\sigma$ \cite{williamson36,simon99}. It can
be shown that, because of the canonical commutation relations, the
positivity of the density matrix $\varrho$ is equivalent to the
following uncertainty relation for the symplectic eigenvalues of the
CM describing a Gaussian state: \be \nu_{j}\ge 1 \, , \quad {\rm
for}\;\; j=1,\ldots,n \, . \label{eigheis} \ee The purity ${\rm
Tr}\,[\varrho^2]$ of a Gaussian state $\varrho$ with CM $\sig$ and
symplectic eigenvalues $\{\nu_{j}\}$ is simply given by \be {\rm
Tr}\,[\varrho^2] = 1/\sqrt{\det{\sig}} =\prod_{j=1}^{n} (1/\nu_j) \,
. \label{purity} \ee The purity quantifies the degree of mixedness
of the Gaussian state $\varrho$, ranging from $1$ for pure states to
the limiting value $0$ for completely mixed states (due to the
infinite dimension of the Hilbert space, no finite lower bound to
the $2$-norm of $\varrho$ exist). Its conjugate $S_{L} = 1 - {\rm
Tr}\,[\varrho^2]$ is referred to as the linear entropy, ranging from
$0$ for pure states to the limiting value $1$ for maximally mixed
states. Another proper way of quantifying the mixedness of a state
is provided by the von Neumann entropy $S_{V} = -\,{\rm
Tr}\,[\varrho\ln \varrho]$. The von Neumann entropy of a Gaussian
state with CM $\sig$ and symplectic eigenvalues $\{\nu_{j}\}$ reads
\cite{serafozzi} \be S_{V} = \sum_{j=1}^{n} f(\nu_j) \; , \ee with
\be f(x)= \frac{x+1}{2}\ln \left(\frac{x+1}{2}\right) -
\frac{x-1}{2}\ln \left(\frac{x-1}{2}\right) \, . \label{entfunc} \ee
Let us now consider a $(m+n)$-mode bipartite Gaussian state {\em
i.e.~}a Gaussian state separated into a subsystem $A$ of $m$ modes,
owned by party $A$, and a subsystem $B$ of $n$ modes, owned by party
$B$. This state is associated to a $2(m+n)$-dimensional CM
$\gr\sigma$. Now, in general, for {\em any} bipartite quantum state
$\varrho$, the positivity of the partially transposed density matrix
$\tilde{\varrho}$, that is, the operator obtained from $\varrho$ by
transposing the variables of only one of the two subsystems, is a
necessary condition for the separability of the state. This
condition thus goes under the name of ``Positivity of Partial
Transpose (PPT) criterion'' \cite{peres96,horodecki96}. This fact is
especially useful when dealing with CV systems, as the action of
partial transposition on CMs can be stated mathematically in very
simple terms: the CM $\tilde{{\gr\sigma}}$ of the partially
transposed state $\tilde{\varrho}$ with respect to, say, subsystem
$A$, is simply obtained by switching the signs of the $m$ momenta
$\{p_{j}\}$ belonging to subsystem $A$ \cite{simon00}: \be
\tilde{{\gr\sigma}} = T {\gr\sigma} T \, , \quad {\rm with} \quad T
\equiv \bigoplus_{1}^{m} \left(\begin{array} {cc}
1&0\\
0&-1
\end{array}\right) \oplus {\mathbbm 1}_{2n} \; ,
\ee where ${\mathbbm 1}_{2n}$ stands for the $2n$-dimensional
identity matrix. Even more remarkably, it has been proven that the
PPT condition is not only necessary, but as well sufficient for the
separability of $(1+n)$-mode Gaussian states \cite{simon00,werwolf}
and of $(m+n)$-mode bisymmetric Gaussian states \cite{unitarily},
thus providing a powerful theoretical tool to detect quantum
entanglement in these relevant classes of states. Let us notice that
the $(1+n)$-mode bipartitions encompass all the possible
bipartitions occurring in three-mode states. In analogy with
\eq{eigheis}, the PPT criterion can be explicitly expressed as a
condition on the symplectic eigenvalues $\{\tilde{\nu}_j\}$ of the
partially transposed CM $\tilde{\gr\sigma}$: \be \tilde\nu_{j} \ge 1
\, , \quad \mbox{for all} \;\; j=1,\ldots,n \; . \label{eigposi} \ee
We finally mention that, in alternative to the PPT criterion, one
can introduce an operational criterion based on a nonlinear map,
that is independent of, and strictly stronger than the PPT condition
\cite{giedkemappa}. In fact, this criterion is necessary and
sufficient for separability of all $(m+n)$-mode Gaussian states of
any $m \times n$ bipartions.

For future convenience, let us define and
write down the CM  $\sig_{1\ldots n}$
of an $n$-mode Gaussian state in terms of two by two
submatrices as
\be
\sig_{1\ldots n} = \left(\begin{array}{cccc}
\sig_{1} & \eps_{12}\; & \cdots & \eps_{1n} \\
&&&\\
\eps_{12}^{\sf T}\; & \ddots & \ddots & \vdots \\
&&&\\
\vdots & \ddots & \ddots & \eps_{n-1 n} \\
&&&\\
\eps_{1n}^{\sf T}& \cdots & \eps_{n-1 n}^{\sf T} & \sig_{n} \\
\end{array}\right) \; . \label{subma}
\ee
The symplectic eigenvalues $\nu_{\mp}$ of a
two-mode CM $\sig_{12}$ are invariant
under symplectic operations acting on $\sig_{12}$.
Starting from this observation, it has been shown that
they can be retrieved from the
knowledge of the symplectic invariants $\det \sig_{12}$ and
$\Delta_{12}=\det \sig_1+\det \sig_2+2 \det \eps_{12}$, according to
the following formula \cite{logneg,serafozzi}:
\be
2\nu_{\mp}^2 = \Delta_{12} \mp
\sqrt{\Delta_{12}^2-4\det \sig_{12}} \; .
\ee
The uncertainty
relation \eq{eigheis} imposes
\be
\Delta_{12} - \det{\sig_{12}}
\le 1 \; .
\label{unce}
\ee
Likewise, the symplectic eigenvalues ${\tilde{\nu}}_{\mp}$
of the CM ${\tilde{\sig}}_{12}$ of the partially transposed
state can be determined by partially
transposing such invariants and can thus be easily computed as
\be
2{\tilde{\nu}}_{\mp}^2 = \tilde\Delta_{12} \mp \sqrt{\tilde\Delta_{12}^2-4\det
\sig_{12}} \; ,
\ee
where $\tilde{\Delta}_{12}=\det
\sig_1+\det\sig_2-2\det\eps_{12}$.

Let us finally observe that the quantities
$$\Delta_{1\ldots n}\equiv \sum_{j=1}^{n}\det\sig_{j} + 2 \sum_{j<k}\det{\eps_{jk}}$$
are symplectic invariants for any number $n$ of modes \cite{serafozzi05}.

We now move on to review in some detail the possible entanglement measures
apt to quantify the entanglement of two-mode Gaussian states, upon
which multipartite counterparts will be constructed in the
following.

\subsection{Quantifying the entanglement of two-mode Gaussian states}

Thanks to the necessary and sufficient PPT criterion for separability,
a proper measure of entanglement for two-mode Gaussian states is
provided by the {\em negativity} $\N$, first introduced in
Ref.~\cite{zircone}, later thoroughly discussed and extended in
Refs.~\cite{logneg,jenstesi} to CV systems. The negativity of a
quantum state $\varrho$ is defined as
\be
{\cal N}(\varrho)=\frac{\|\tilde \varrho \|_1-1}{2}\: ,
\ee
where $\tilde\varrho$ is the partially transposed density matrix and
$\|\hat o\|_1=\,{\rm Tr}|\hat o|$ stands for the trace norm of the
hermitian operator $\hat o$.  This measures quantifies the extent to
which $\tilde\varrho$ fails to be positive. Strictly related to $\N$
is the {\em logarithmic negativity} $E_{\N}$, defined as
$E_{\N}\equiv \ln \|\tilde{\varrho}\|_{1}$, which constitutes an
upper bound to the {\em distillable entanglement} of the quantum
state $\varrho$ and is related to the entanglement cost under PPT
preserving operations \cite{auden03}. Both the negativity and the
logarithmic negativity have been proven to be monotone under LOCC
(local operations and classical communication)
\cite{logneg,jenstesi,plenio05}, a crucial property for a {\em bona
fide} measure of entanglement. Moreover, the logarithmic negativity
possesses the agreeable property of being additive. For any
two--mode Gaussian state $\varrho$ it is easy to show that both the
negativity and the logarithmic negativity are simple decreasing
functions of the lowest symplectic eigenvalue
$\tilde{\nu}_{-}$ of the CM of the partially transposed
state \cite{logneg,extremal}:
\be
\|\tilde\varrho\|_{1}=\frac{1}{\tilde{\nu}_{-}}\;\Rightarrow
\N(\varrho)=\max \, \left[
0,\frac{1-\tilde{\nu}_{-}}{2\tilde{\nu}_{-}} \right] \, , \ee \be
\label{ennu} E_{\N}(\varrho)=\max\,\left[0,-\ln
\tilde{\nu}_{-}\right] \, .
\ee
These expressions directly quantify
the amount by which the necessary and sufficient PPT condition
(\ref{eigposi}) for separability is violated. The lowest symplectic
eigenvalue $\tilde{\nu}_{-}$ of the partially transposed
state ${\tilde{\sig}}$ thus completely qualifies and
quantifies, in terms of negativities, the entanglement of a
two--mode Gaussian state $\sig$. For $\tilde \nu_- \ge 1$ the state
is separable, otherwise it is entangled; moreover, in the limit of
vanishing $\tilde \nu_-$, the negativities, and thus the entanglement,
diverge.

In the special instance of symmetric two--mode Gaussian states ({\em
i.e.}~of states with $\det\sig_1=\det\sig_2$), the {\em entanglement
of formation} (EoF) \cite{bennet96}, can be computed as well
\cite{eofprl}. We recall that the EoF $E_{F}$ of a quantum state
$\varrho$ is defined as
\be
E_{F}(\varrho)=\min_{\{p_i,\ket{\psi_i}\}}\sum_i p_i E(\ket{\psi_i})
\; ,
\label{eof}
\ee
where $E(\ket{\psi_i})$ denotes the von Neumann entropy $S_V$
of the reduced density matrix of one party in the pure states
$E(\ket{\psi_i})$, namely the unique measure
of bipartite entanglement for all pure quantum states (entropy of
entanglement). The minimum in \eq{eof} is taken over all the pure
states realizations of $\varrho$:
\[
\varrho=\sum_i p_i \ket{\psi_i}\bra{\psi_i} \; .
\]
The asymptotic regularization of the entanglement of formation
coincides with the {\em entanglement cost} $E_C (\varrho)$, defined
as the minimum number of singlets (maximally entangled antisymmetric
two-qubit states) which is needed to prepare the state $\varrho$
through LOCC \cite{ecost}.

The optimal convex decomposition of \eq{eof} has been determined
exactly for symmetric two--mode Gaussian states, and turns out to
be Gaussian, that is, the absolute minimum is realized within the
set of pure two--mode Gaussian states, yielding \cite{eofprl}
\be
E_F =
\max\left[ 0,h(\tilde{\nu}_{-}) \right] \; ,
\label{eofgau}
\ee
with
\be
\label{hentro}
h(x)=\frac{(1+x)^2}{4x}\ln \left[\frac{(1+x)^2}{4x}\right]-
\frac{(1-x)^2}{4x}\ln \left[\frac{(1-x)^2}{4x}\right].
\ee
Such a quantity is, again, a monotonically decreasing function of
$\tilde{\nu}_{-}$. Therefore it provides a quantification of the
entanglement of symmetric states {\em equivalent} to the one
provided by the negativities. This equivalence, regrettably, does
not hold for general, mixed nonsymmetric states. In this case
the EoF is not computable; nonetheless, it has been demonstrated
that different entanglement measures induce different orderings of the
states \cite{ordering}. This means that, depending on the measure
of entanglement that one chooses, either the PPT-inspired
negativities or the entropy-based Gaussian measures (see below),
a certain state can be more or less entangled than another
given state. Clearly, this is neither a catastrophic nor
an entirely unexpected result, but rather a consequence of the fact
that, in general, for mixed states, different measures of
entanglement may be associated to different conceptual and
operational definitions, and thus may measure different aspects
of the quantum correlations present in a statistical mixture.

In fact, restricting to the Gaussian framework, a special family of
proper entanglement measures can be defined, sharing the agreeable
property of being analytically computable in several instances of
physical interest. The formalism of {\em Gaussian entanglement
measures} (Gaussian EMs), first introduced in Ref. \cite{geof}, has
been further developed and analysed in Ref.~\cite{ordering}. Such a
formalism enables to define generic Gaussian EMs of bipartite
entanglement by applying the Gaussian convex roof, that is, the
convex roof over pure Gaussian decompositions only, to any {\em bona
fide} measure of bipartite entanglement defined for pure Gaussian
states. As already mentioned, the optimization problem \eq{eof} for
the computation of the EoF of nonsymmetric two--mode Gaussian states
has not yet been solved. However, the task can be somehow simplified
by restricting to decompositions into pure Gaussian states only. The
resulting measure, named ``Gaussian EoF'' in Ref. \cite{geof}, is an
upper bound to the true EoF and coincides with it for symmetric
two--mode Gaussian states.

In general, we can define a Gaussian EM $G_E$ as follows. For any
pure Gaussian state $|\psi \rangle $ with CM $\sig^P$, one has
\begin{equation}
\label{Gaussian EMp}
G_E (\sig^P) \equiv E(|\psi \rangle ) \; ,
\end{equation}
where $E$ can be {\em any} proper measure of entanglement of pure
states, defined as a monotonically increasing function of the
entropy of entanglement ({\ie}the von Neumann entropy of the reduced
density matrix of one party).

For any mixed Gaussian state $\varrho$ with CM $\sig$, one has
\cite{geof}
\begin{equation}
\label{Gaussian EMm}
G_E (\sig) \equiv \inf_{\sig^P \le \sig} G_E(\sig^P) \; .
\end{equation}
If the function $E$ is taken to be exactly the entropy of
entanglement, then the corresponding Gaussian EM defines
the Gaussian entanglement of formation
({\em Gaussian EoF}) \cite{geof}.
From an operational point of view, the Gaussian EoF is strictly
related to the capacity of bosonic Gaussian channels
\cite{jensata}. Moreover, the Gaussian EoF is an entanglement
monotone under Gaussian LOCC, a property that is shared by all
Gaussian EMs \cite{geof,ordering}.

In general, the definition \eq{Gaussian EMm} involves an
optimization over all pure Gaussian states with CM $\sig^P$ smaller
than the CM $\sig$ of the mixed state whose entanglement one wishes
to compute. This is a simpler optimization problem than that
appearing in the definition \eq{eof} of the true EoF, which, in CV
systems, would imply considering decompositions over all, Gaussian
{\it and} non-Gaussian pure states. Despite this simplification,
in general the Gaussian EMs cannot be expressed in a simple closed
form, even for two--mode Gaussian states. However, the Gaussian
EMs have been computed analytically \cite{ordering} for two relevant
classes of, {\it generally nonsymmetric}, two--mode Gaussian states,
namely the states of {\em extremal} -- maximal and minimal --
negativity at fixed global and local purities, referred to,
respectively as Gaussian Maximally Entangled Mixed States (GMEMS)
and Gaussian Least Entangled Mixed States (GLEMS) \cite{prl,extremal}. In
particular, the explicit expression of the Gaussian EMs of the GLEMS will be
crucial in the following because, as we are anout to show, any
two--mode reduction of a three--mode pure Gaussian state is a GLEM.

\section{Three-mode Gaussian states}\label{secbarbie}

To begin with, let us set the notation and review the known
results about three-mode Gaussian states of CV systems.
We will refer to the three modes under exam as mode $1$, $2$ and $3$.
The two by two submatrices that form the CM $\sig \equiv \sig_{123}$
of a three-mode Gaussian state are defined according to \eq{subma},
whereas the four by four CMs of the reduced two-mode Gaussian states
of modes $i$ and $j$ will be denoted by $\sig_{ij}$.
Likewise, the local symplectic invariants $\Delta_{ij}$ will be
specified by the labels $i$ and $j$ of the modes they refer to,
while, to avoid any confusion, the three-mode (global) symplectic
invariant will be denoted by $\Delta\equiv\Delta_{123}$. Let us
recall the uncertainty relation \eq{unce} for two-mode
Gaussian states:
\be
\Delta_{ij} - \det{\sig_{ij}} \le 1 \; .
\label{uncedue}
\ee
As we have seen in the previous section, a complete {\em
qualitative} characterization of the entanglement of three-mode
Gaussian state is possible because the PPT criterion is
necessary and sufficient for their separability under {\em any}, partial
or global, bipartition. This has lead to an exhaustive
classification of three-mode Gaussian states in five distinct
classes \cite{kraus}. These classes take into account the fact that
the modes $1$, $2$ and $3$ allow for three distinct bipartitions:
\begin{itemize}
\item{Class 1: states not separable under all the three possible
bipartitions $i \times (jk)$  of the modes (fully inseparable states,
possessing genuine multipartite entanglement).}
\item{Class 2: states separable under only one of the three possible
bipartitions (one-mode biseparable states).}
\item{Class 3: states separable under only two of the three possible
bipartitions (two-mode biseparable states).}
\item{Class 4: states separable under all the three possible bipartitions,
but impossible to write as a convex sum of tripartite products of pure
one-mode states (three-mode biseparable states).}
\item{Class 5: states that are separable under all the three possible bipartitions,
and can be written as a convex sum of tripartite
products of pure one-mode states (fully separable states).}
\end{itemize}
Notice that classes 4 and 5 cannot be distinguished by partial
transposition of any of the three modes (which is positive for both
classes). States in class 4 stand therefore as nontrivial
examples of tripartite entangled states of CV systems
with positive partial transpose \cite{kraus}. It is well known
that entangled states with positive partial transpose possess
{\it bound entanglement}, that is, entanglement that cannot be
distilled by means of LOCC.

\subsection{Pure states}\label{secpuri}

We begin by focusing on {\em pure} three-mode Gaussian states, for
which one has
\be
\det{\sig} = 1 \; , \quad \Delta=3 \, .
\label{purinv}
\ee
The purity constraint requires the
local entropic measures of any $1\times 2$-mode bipartitions to be
equal:
\be
\det{\sig_{ij}}=\det{\sig_{k}} \; ,
\label{pur}
\ee
with $i$, $j$ and $k$ different from each other.
This general, well known property of the bipartitions of pure
states may be easily proven resorting to the Schmidt decomposition.

A first consequence of Eqs.~(\ref{purinv}) and (\ref{pur}) is rather
remarkable. Combining such equations one easily obtains
\begin{eqnarray}
(\Delta_{12}-\det{\sig_{12}}) & + & (\Delta_{13}-\det{\sig_{13}}) \nonumber \\
& + & (\Delta_{23}-\det{\sig_{23}}) = 3 \; ,
\end{eqnarray}
which, together with Inequality (\ref{uncedue}), implies \be
\Delta_{ij} = \det{\sig_{ij}} + 1 \; , \quad \forall \, i,j: \;
i\neq j \, . \label{glems} \ee The last equation shows that any
reduced two-mode state of a pure three-mode Gaussian state saturates
the partial uncertainty relation \eq{uncedue}. The states endowed
with such a partial minimal uncertainty are states of minimal
negativity for given global and local purities, Gaussian least
entangled mixed states (GLEMS) \cite{prl,extremal}. We recall that
by two-mode mixed Gaussian states of partial minimum Heisenberg
uncertainty one means states that have one of the two symplectic
eigenvalues equal to $1$. States with both symplectic eigenvalues
equal to $1$ are of course the pure Gaussian states of absolute
minimum Heisenberg uncertainty. These definitions immediately extend
to arbitary multimode Gaussian states. In this case, $n$-mode
Gaussian states of $m$-partial minimum uncertainty are those that
have $m$ out of the $n$ symplectic eigenvalues equal to $1$, with
$m<n$. Notice that such a result could have also been inferred by
invoking the reduction of $(1+n)$-mode pure Gaussian states
discussed in Ref.~\cite{botero}, first introduced in
Ref.~\cite{holevowerner} and proved at the covariance matrix level
in Ref.~\cite{giedke03}. This implies that, through local unitaries
(under any bipartition of the three modes), the state can be brought
to the product of a two-mode squeezed state and of an uncorrelated
vacuum. In turn, this implies that any of the three reduced two-mode
CMs (resulting from the discarding of one mode) has one symplectic
eigenvalue equal to $1$ and is thus a GLEM.

In fact, our simple proof, straightforwardly derived in terms of
symplectic invariants, provides some further insight into the
structure of CMs characterizing Gaussian states. What matters to our
aims, is that the standard form CM of Gaussian states is completely
determined by their global and local invariants. Therefore, because
of \eq{pur}, the entanglement between any pair of modes embedded in
a three-mode pure Gaussian state is fully determined by the local
invariants $\det{\sig_{l}}$, for $l=1,2,3$, whatever proper measure
we choose to quantify it \cite{ordering}. Furthermore, the
entanglement of a $\sig_i|\sig_{jk}$ bipartition of a pure
three-mode state is determined by the entropy of one of the reduced
states that is, once again, by the quantity $\det{\sig_{i}}$. Thus,
{\em the three local symplectic invariants $\det{\sig_{1}}$,
$\det{\sig_{2}}$ and $\det{\sig_{3}}$ fully determine the
entanglement of any bipartition of a pure three-mode Gaussian
state}. We will show that they suffice to determine as well the
genuine tripartite entanglement encoded in the state.

For ease of notation, in the following we will denote by $a_l$ the
local single-mode symplectic eigenvalues associated to mode $l$ with
CM $\sig_l$: \be \label{al} a_l\equiv \sqrt{\det{\sig_l}} \; . \ee
\eq{purity} shows that the quantities $a_l$ are simply related to
the purities of the reduced single-mode states, the local purities
$\mu_l$, by the relation \be \mu_l = \frac{1}{a_{l}} \; . \ee Since
the set $\{a_l\}$ fully determines the entanglement of any of the
$1\times2$--mode and $1\times1$--mode bipartitions of the state, it
is important to determine the range of the allowed values for such
quantities. This will provide a complete quantitative
characterization of the entanglement of three-mode pure Gaussian
states. To this aim, let us focus on the reduced two-mode CM
$\sig_{12}$ and let us bring it (by local unitaries) in standard
form \cite{duan00,simon00}, so that \eq{subma} is recast in the form
\begin{eqnarray}
\label{2sform}
\sig_{l} & = & {\rm diag}\{a_{l},\,a_{l}\}\,,\quad
l=1,2 \, ; \nonumber \\
\eps_{12} & = & {\rm diag}\{c_{12},\,d_{12}\} \; ,
\end{eqnarray}
where $c_{12}$ and $d_{12}$ are the two-mode covariances, and, as we
will show below, can be evaluated independently in pure three-mode
Gaussian states. Notice that no generality is lost in assuming a
standard form CM, because the entanglement properties of any
bipartition of the system are invariant under local (single-mode)
symplectic operations. Now, Eqs.~(\ref{pur}) and (\ref{purinv}) may
be recast as follows \bea
a_3^2 &=& a_1^2 + a_2^2 +2 c_{12} d_{12} -1 \; , \\
a_3^2 &=& (a_1 a_2 -c_{12}^2)(a_1 a_2 -d_{12}^2) \; , \eea showing
that we may eliminate one of the two covariances to find the
expression of the remaining one only in terms of the three local
inverse of the purities $a_l$ (mixednesses). Defining the quantity
$\kappa$ as \be \kappa\equiv c_{12} d_{12} =
\frac{1+a_3^2-a_1^2-a_2^2}{2} \; , \ee leads to the following
condition on the covariance $c_{12}$: \be c_{12}^4 - \frac{1}{a_1
a_2}\left[(\kappa-1)^2+a_1^2 a_2^2-a_1^2-a_2^2\right] c_{12}^2 +
\kappa^2 = 0 \; . \label{eqe1} \ee Such a second order algebraic
equation for $c_{12}^2$ admits a positive solution if and only if
its discriminant $\delta$ is positive: \be \delta \ge 0 \; .
\label{deltino} \ee After some algebra, one finds \bea
\delta&=&(a_1+a_2+a_3+1)(a_1+a_2+a_3-1) \nonumber\\
&\times& (a_1+a_2-a_3+1)(a_1-a_2+a_3+1) \nonumber \\
&\times&(-a_1+a_2+a_3+1)(a_1+a_2-a_3-1) \nonumber \\
&\times&(a_1-a_2+a_3-1)(-a_1+a_2+a_3-1) \, .
\eea
Aside from the existence of a real covariance $c_{12}$, the further
condition of positivity of $\sig_{12}$ has to be fulfilled for a
state to be physical. This amounts to impose the inequality $a_1 a_2
-c_{12}^2 \ge 0$, which can be explicitly written, after solving
\eq{eqe1}, as
$$
4\left[2a_1^2 a_2^2 -
\left((\kappa-1)^2+a_1^2a_2^2-a_1^2-a_2^2\right)\right] \ge
\sqrt{\delta} \; .
$$
This inequality is trivially satisfied when squared on both sides;
therefore it reduces to
\be
2a_1^2a_2^2 - \left((\kappa-1)^2+a_1^2a_2^2-a_1^2-a_2^2\right)\ge 0
\; .
\label{posi}
\ee

\begin{figure*}[t!]
\centering{\includegraphics[width=14cm]{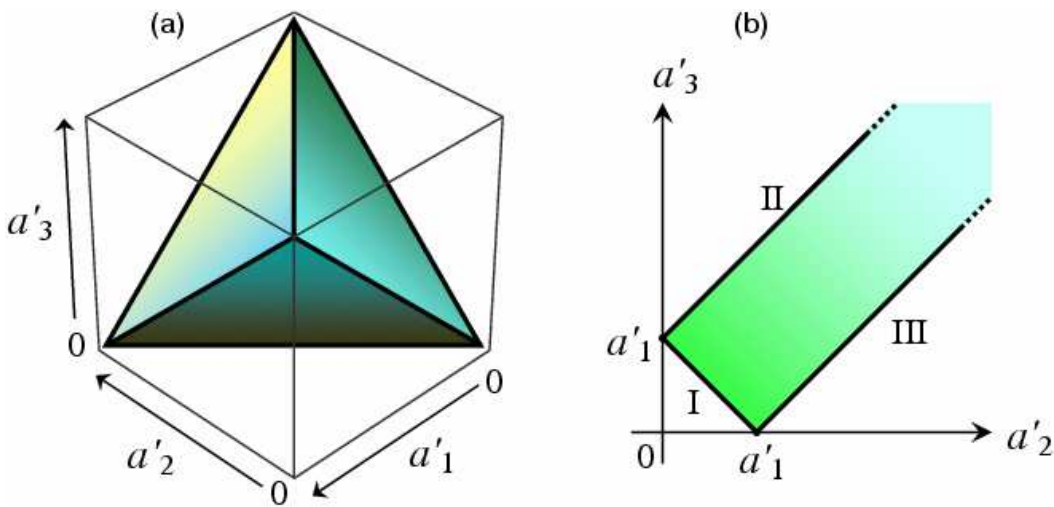}%
\caption{\label{figangle}(color online) Range of the entropic
quantities $a'_l = \mu_l^{-1}-1$  for pure three-mode Gaussian
states. The three parameters $a'_l$, with $l=1,2,3$, have to vary
inside the pyramid represented in plot \textsf{(a)} or,
equivalently, for fixed values of one of them, say $a'_1$, inside
the shaded slice represented in plot \textsf{(b)}, in order to
determine the CM of a physical state, \eq{cm3tutta}. The expression
of the boundary surfaces/curves come from the saturation of the
triangular inequality (\ref{triangleprim}) for all possible mode
permutations. In particular, for the projected two-dimensional plot
\textsf{(b)}, the equations of the three boundaries are:
I.~$a'_3=a'_1-a'_2$; II.~$a'_3=a'_1+a'_2$; III.~$a'_3=a'_2-a'_1$.}}
\end{figure*}

Notice that conditions (\ref{deltino}) and
(\ref{posi}), although derived by assuming a specific bipartition of
the three modes, are independent on the choice of the modes
that enter in the considered bipartition, because they are
invariant under all possible permutations of the modes.
Defining the parameters
\begin{equation}
\label{aprimi}
a'_l \equiv a_l-1 \; ,
\end{equation}
the Heisenberg uncertainty principle for single-mode states reduces
to \be a'_l \ge 0 \; \; \; \; \; \forall l=1,2,3 \; . \ee This fact
allows to greatly simplify the two previous conditions, which can be
combined into the following triangular inequality \be
\label{triangleprim} |a'_i-a'_j| \, \le \, a'_k \, \le \, a'_i+a'_j
\; . \ee Inequality (\ref{triangleprim}) is a condition invariant
under all possible permutations of the mode indexes $\{i,j,k\}$,
and, together with the positivity of each $a'_l$, fully
characterizes the local symplectic eigenvalues of the CM of
three-mode pure Gaussian states. It therefore provides a complete
characterization of the entanglement in such states. All standard
forms of pure three-mode Gaussian states and in particular,
remarkably, all the possible values of the logarithmic negativity
between {\em any} pair of subsystems, can be determined by letting
$a'_1$, $a'_2$ and $a'_3$ vary in their range of allowed values, as
summarized in Fig. \ref{figangle}.

Let us remark that \eq{triangleprim} qualifies itself as an entropic
inequality, as the quantities $\{a'_j\}$ are closely related to the
purities and to the von Neumann entropies of the single-mode reduced
states. In particular the von Neumann entropies $S_{Vj}$ of the
reduced states are given by $S_{Vj}=f(a'_j+1)=f(a_j)$, where the
increasing convex entropic function $f(x)$ has been defined in
\eq{entfunc}. Now, Inequality (\ref{triangleprim}) is strikingly
analogous to the well known triangle (Araki-Lieb) and subadditivity
inequalities for the von Neumann entropy (holding for general
systems, see, {\em e.g.}, \cite{chuaniels}), which in our case read
\be |f(a_i)-f(a_j)| \le f(a_k) \le f(a_i) + f(a_j) \; .
\label{arakilieb} \ee However, as the different convexity properties
of the functions involved suggest, Inequalities (\ref{triangleprim})
and (\ref{arakilieb}) are not equivalent. Actually, as can be shown
by exploiting the properties of the function $f(x)$, the
Inequalities (\ref{triangleprim}) imply the Inequalities
(\ref{arakilieb}) for both the leftmost and the rightmost parts. On
the other hand, there exist values of the local symplectic
eigenvalues $\{a_j\}$ for which Inequalities (\ref{arakilieb}) are
satisfied but (\ref{triangleprim}) are violated. Therefore, the
condition imposed by \eq{triangleprim} is stronger than the
generally holding inequalities for the von Neumann entropy applied
to pure states.

Let us recall that the form of the CM of any Gaussian state can be
simplified through local (unitary) symplectic operations (that
therefore do not affect the entanglement or mixedness properties
of the state) belonging to $Sp_{2,\R}^{\oplus n}$.
Such reductions of the CMs are called ``standard forms''.
For the sake of clarity, let us write the explicit standard form
CM of a generic {\em pure} three-mode Gaussian state:
\begin{equation}
\label{cm3tutta}
\sig^p_{sf}=\left(
\begin{array}{cccccc}
 a_1 & 0 & e_{12}^+ & 0 & e_{13}^+ & 0 \\
 0 & a_1 & 0 & e_{12}^- & 0 & e_{13}^- \\
 e_{12}^+ & 0 & a_2 & 0 & e_{23}^+ & 0 \\
 0 & e_{12}^- & 0 & a_2 & 0 & e_{23}^- \\
 e_{13}^+ & 0 & e_{23}^+ & 0 & a_3 & 0 \\
 0 & e_{13}^- & 0 & e_{23}^- & 0 & a_3
\end{array}
\right)\; ,
\end{equation}
with
\begin{widetext}
\begin{equation}\label{eij}
e_{ij}^{\pm} \equiv
\frac{\sqrt{\left[\left(a_i-a_j\right)^2-\left(a_k-1\right)^2\right]
\left[\left(a_i-a_j\right)^2-\left(a_k+1\right)^2\right]} \pm
\sqrt{\left[\left(a_i+a_j\right)^2-\left(a_k-1\right)^2\right]
\left[\left(a_i+a_j\right)^2-\left(a_k+1\right)^2\right]}}{4
\sqrt{a_i a_j}} \, .
\end{equation}
\end{widetext}
By direct comparison with Eq. (67) in Ref. \cite{extremal}, it is
immediate to verify that each two-mode reduced CM $\sig_{ij}$
denotes a standard form GLEMS with local purities $\mu_i=a_i^{-1}$
and $\mu_j=a_j^{-1}$, and global purity $\mu_{ij}\equiv \mu_k =
a_k^{-1}$. From our study it then turns out that, regarding the
classification of Sec. \ref{secbarbie} \cite{kraus}, pure three-mode
Gaussian states may belong either to class 5, in which case they
reduce to the global three-mode vacuum, or to class 2, reducing to
the uncorrelated product of a single-mode vacuum and of a two-mode
squeezed state, or to class 1 (fully inseparable state). No two-mode
or three-mode biseparable pure three-mode Gaussian states are
allowed.

Let us finally stress that, although useful in actual calculations,
the use of CMs in standard form does not entail any loss of
generality, because all the results derived in the present work
do not depend on the choice of the specific form of the CMs, but
only on invariant quantities, such as the global and local
symplectic invariants.

\subsection{Mixed states}
The most general standard form $\sig_{sf}$ associated to the CM of
any (generally mixed) three-mode Gaussian state can be written as
\begin{equation}\label{tutta}
\sig_{sf}=\left(
\begin{array}{cccccc}
 a_1 & 0 & f_1 & 0 & f_3 & f_5 \\
 0 & a_1 & 0 & f_2 & 0 & f_4 \\
 f_1 & 0 & a_2 & 0 & f_6 & f_8 \\
 0 & f_2 & 0 & a_2 & f_9 & f_7 \\
 f_3 & 0 & f_6 & f_9 & a_3 & 0 \\
 f_5 & f_4  & f_8 & f_7 & 0 & a_3
\end{array}
\right)\, ,
\end{equation}
where the 12 parameters $\{a_j\}$ (inverse of the local purities)
and $\{f_j\}$ (the covariances describing correlations between the
modes) are only constrained by the Heisenberg uncertainty relations
\eq{eigheis}. The possibility of this useful, general reduction can
be easily proven along the same lines as the two-mode standard form
reduction \cite{duan00}: by means of three local symplectic
operations one can bring the three blocks $\sig_1$, $\sig_2$ and
$\sig_3$ in Williamson form, thus making them insensitive to further
local rotations (which are symplectic operations); exploiting such
rotations on mode $1$ and $2$ one can then diagonalize the block
$\gr{\varepsilon}_{12}$ as allowed by its singular value
decomposition; finally, one local rotation on mode $3$ is left, by
which one can cancel one entry of the block $\gr{\varepsilon}_{13}$.
Indeed, the resulting number of free parameters could have been
inferred by subtracting the number of parameters of an element of
$Sp_{2,\R}\oplus Sp_{2,\R}\oplus Sp_{2,\R}$ (which is $9$, as
$Sp_{2,\R}$ has $3$ independent generators) from the 21 entries of a
generic $6\times 6$ symmetric matrix.

\subsection{Symmetric states}\label{secsym}
Among generic Gaussian states, those endowed with some properties
of symmetry under mode exchange play a special role for what concerns
the structure of entanglement. In particular, in a three-mode CV system,
{\em bisymmetric} states are Gaussian states
invariant under the exchange of two given modes (say $2$ and $3$)
\cite{unitarily,adescaling}.
Their CM will be thus of the form
\begin{equation}
\label{biscm}
\sig_{bis}={\left(%
\begin{array}{ccc}
\gr\alpha & \gr\varepsilon &  \gr\varepsilon \\
\gr\varepsilon^{\sf T} & \gr\beta & \gr\zeta\\
\gr\varepsilon^{\sf T}  & \gr\zeta^{\sf T} & \gr\beta \\
\end{array}%
\right)}\,.
\end{equation}
Let mode $1$ be entangled with the block of modes $(23)$. It has
been proven \cite{adescaling,unitarily} that for such bisymmetric
states the application of a {\em local unitary} (symplectic in phase
space) operation on the block $(23)$ concentrates the whole original
multimode entanglement into the reduced state of a single pair of modes.
Namely, in terms of the new modes $\{1, 2',3'\}$, the CM is
transformed in a two-mode entangled state of modes $1$ and $2'$,
tensor the uncorrelated single-mode state of mode $3'$, so that the original
multimode entanglement can be quantified resorting to the well established
theory of bipartite entanglement in two-mode Gaussian states
\cite{prl,extremal,eisplenio,ciracfortshit,francamentemeneinfischio}.

The local symplectic transformation responsible for the unitary
localization of the multimode entanglement is typically realized by
a simple beam splitter, if the CM is in standard form, with the
single-mode blocks in their Williamson diagonal form. More
generally, it may be a combination of beam splitters, phase shifters
and squeezers. This type of entanglement localization is unitary and
reversible, and thus completely different from the usual
localization or concentration procedures that are based on
measurements, as in the case of the ``localizable entanglement''
previously introduced for spin systems
\cite{localizprl,localizpra}). To reconstruct the original state, it
suffices to let the discarded mode $3'$ interfere once more with
mode $2'$ through the reversed beam splitter (that is, by applying
the inverse symplectic operation). We remark that the unitary
localizability is a property that extends to all $1 \times n$
Gaussian states \cite{adescaling}, and to all $m \times n$
 bisymmetric Gaussian states \cite{unitarily}, enabling two parties
(owing two respective blocks of multiple symmetric modes) to
realize, by purely local controls, a perfect and reversible
entanglement switch between two-mode and multimode quantum
correlations.

Three-mode Gaussian states which are invariant under the exchange of {\em
any} two modes are said to be {\em fully symmetric}. They are
trivially bisymmetric with respect to any $1 \times 2$ bipartition,
meaning that each conceivable bipartite entanglement is locally
equivalent to two-mode entanglement. In the Gaussian setting, these
states are described by a CM \cite{adescaling,unitarily}
\begin{equation}
\label{fscm}
\sig_{s}={\left(%
\begin{array}{ccc}
\gr\alpha & \gr\varepsilon &  \gr\varepsilon \\
\gr\varepsilon^{\sf T} & \gr\alpha & \gr\varepsilon\\
\gr\varepsilon^{\sf T}  & \gr\varepsilon^{\sf T} & \gr\alpha \\
\end{array}%
\right)}\,,
\end{equation}
where the local mixedness $a \equiv \sqrt{\det{\gr\alpha}}$ is the
same for all the three modes. These states have been successfully
produced in laboratory by quantum optical means
\cite{3mexp,pfister}, and exploited to implement quantum
teleportation networks \cite{network,naturusawa}. Used as
shared resources, they can be optimized with respect to
local operations to realize CV teleportation with
maximal nonclassical fidelity \cite{telepoppate},
quantum secret sharing \cite{secret}, controlled dense coding
\cite{dense}, and to solve CV Byzantine agreement \cite{sanpera}.
Moreover, the structure of tripartite entanglement in this kind of
states presents peculiar sharing properties \cite{contangle}, that
are quite different from the properties of distributed entanglement
among qubits and qudits \cite{sharing}, as will be discussed in
detail in Sec. \ref{secpromis}.

We finally mention that the unitary localizability of entanglement
does not apply only to states with special symmetries. For instance,
for all {\em pure} three-mode Gaussian states, the $1 \times 2$
entanglement can be unitarily localized in any bipartition. This
fact holds for generic pure Gaussian states of $1 \times n$
bipartitions. \cite{botero,unitarily,adebook}.

\section{Genuine tripartite entanglement and entanglement sharing}

In this section we approach in a systematic way the question of
distributing quantum correlations among three parties globally
prepared in a (pure or mixed) three-mode Gaussian state, and we deal
with the related problem of quantifying genuine tripartite
entanglement in such a state.

\subsection{Entanglement sharing}

The key ingredient of our analysis is the so-called {\em sharing} or
{\em monogamy} inequality, first introduced by Coffman, Kundu, and
Wootters (CKW) \cite{CKW} for systems of three qubits, and recently
extended to systems of $n$ qubits by Osborne and Verstraete \cite{Osborne}.
The CKW monogamy inequality for a three-party system can
be written as follows:
\begin{equation}
\label{CKWine}
E^{i|(jk)}-E^{i|j}-E^{i|k} \ge 0 \; ,
\end{equation}
where $i,\,j,\,k$ denote the three elementary parties (modes in a CV
system), and $E$ refers to a proper measure of bipartite
entanglement (in particular, nonnegative on inseparable states and
monotonic under LOCC).

It is natural to expect that \ineq{CKWine} should hold for states of
CV systems as well, despite the fact that they are defined on
infinite-dimensional Hilbert spaces and can in principle achieve
infinite entanglement, in particular the entanglement of
distillation can become infinite in certain states of CV systems;
these states can be defined and constructed rigorously using the
techniques of field theory and statistical mechanics for the
description of systems of infinitely many degrees of freedom
\cite{Keyl}. In fact, one can show that the linearity of quantum
mechanics, through the so-called no-cloning theorem
\cite{nocloning0,nocloning1,nocloning2}, prevents quantum
correlations from being freely shareable, at striking variance with
the behaviour of classical correlations \cite{sharing}. This entails
that quantum entanglement is ``monogamous'' \cite{monogam}.

The crucial issue in contructing and proving the CV version of the
CKW monogamy inequality is to find a proper measure of entanglement
$E$, able to capture the trade-off between couplewise and tripartite
correlations, quantitatively formalized by \ineq{CKWine}. For qubit
systems, such a measure is known as the {\em tangle} \cite{CKW}. For
Gaussian states of CV systems, this problem has been recently solved
in Ref. \cite{contangle}, where the CV analogue of the tangle has been
defined and exploited to obtain a proof of the monogamy inequality
(\ref{CKWine}) for all Gaussian states of three modes, and for all
symmetric Gaussian states of systems with an arbitrary number of modes.
Following the approach of Ref.~\cite{contangle}, we recall now the
notation leading to the definition of the {\em continous-variable
tangle}, and provide a detailed proof of the CKW monogamy inequality
obeyed by all three-mode Gaussian states.

\subsection{The continuous-variable tangle}

The continuous-variable tangle $E_\tau$ is formally defined as
follows \cite{contangle}. For a generic pure state $\ket{\psi}$ of a
$(1+N)$--mode CV system, one has
\begin{equation}
\label{etaupure}
E_\tau (\psi) \equiv \ln^2 \| \tilde \rho \|_1 \; ,\qquad \rho =
\ketbra\psi\psi \; .
\end{equation}
This is a proper measure of bipartite entanglement, being a convex,
increasing function of the logarithmic negativity $E_\N$,
equivalent to the entropy of entanglement on pure states.
For a pure Gaussian state $\ket\psi$ with CM $\sig^p$, it is
easy to find that
\begin{equation}\label{etaupgau}
E_\tau (\sig^p) = \arcsinh^2
\left(\frac{\sqrt{1-\mu_1^2}}{\mu_1}\right) \; ,
\end{equation}
where $\mu_1 = 1/\sqrt{\det\sig_1}$ is the local purity of the
reduced state of mode $1$, described by a CM $\sig_1$ (we are
considering a most general $1 \times n$ bipartition).
Def.~(\ref{etaupure}) is naturally extended to generic mixed states
$\rho$ of $(n+1)$--mode CV systems through the convex-roof formalism
\cite{osbcroof}. Namely,
\begin{equation}
\label{etaumix}
E_\tau(\rho) \equiv \inf_{\{p_i,\psi_i\}} \sum_i p_i
E_\tau(\psi_i) \, ,
\end{equation}
where the infimum is taken over {\em all} convex decompositions of
$\rho$ in terms of pure states $\{\ket{\psi_i}\}$.
If the index $i$ is continuous, the sum in \eq{etaumix}
is replaced by an integral, and the probabilities $\{p_i\}$ by a
probability distribution $\pi(\psi)$.

Next, it is important to recall that for two qubits the tangle can
be equivalently defined as the convex roof of the squared negativity
\cite{Lee}, because the latter coincides with the concurrence for
pure two-qubit states \cite{Wootters}. Then, \eq{etaumix} states
that the convex roof of the squared logarithmic negativity defines
the proper continuous-variable tangle, or, in short, the {\em
contangle} $E_\tau (\rho)$ \cite{contangle}. One could  have defined
the contangle using the convex roof extension of the squared
negativity as well. The two definitions are, in fact, equivalent to
the aim of quantifying distributed entanglement, because the squared
negativity is a convex function of the squared logarithmic
negativity \cite{CKW,sharing}. The nice feature of using
specifically the squared logarithmic negativity lies in the fact
that from a computational point of view the logarithm accounts in a
straightforard way for the infinite dimensionality of the underlying
Hilbert sapce \cite{contangle}. We will prove in the following that
the contangle satisfies the CKW monogamy inequality for all
three-mode Gaussian states. Viceversa, one can easily show that any
continuous-variable tangle defined in terms of the (not squared)
negativity or of the entanglement of formation fails to satisfy the
CKW monogamy inequality  in general \cite{contangle}. This situation
is to some extent reminiscent of the case of qubit systems, for
which the CKW monogamy inequality holds using the tangle, defined as
the convex roof of the squared concurrence \cite{CKW} or of the
squared negativity \cite{Lee}, but fails if one chooses alternative
definitions based on the convex roof of other equivalent measures of
bipartite entanglement, such as the concurrence itself or the
entanglement of formation \cite{CKW}.

From now on, we restrict our attention to Gaussian states. Any
multimode mixed Gaussian state with CM $\sig$, admits a
decomposition in terms of pure Gaussian states only. The infimum of
the average contangle, taken over all pure {\em Gaussian} state
decompositions, defines the {\it Gaussian contangle} $G_\tau$
\begin{equation}
G_\tau(\sig) \equiv \inf_{\{\pi(d\sig^p ), \sig^{p} \}} \int \pi
(d\sig^p) E_\tau (\sig^p) \; .
\label{GaCoRo}
\end{equation}
It follows from the convex roof construction that the Gaussian
contangle $G_\tau(\sig)$ is an upper bound to the true
contangle $E_\tau(\sig)$ (because the latter can be
in principle minimized over a non-Gaussian decomposition):
\begin{equation}
E_\tau(\sig) \, \leq \, G_\tau(\sig) \; ,
\label{uppercut}
\end{equation}
and it can be shown that $G_\tau(\sig)$ is a bipartite entanglement
monotone under Gaussian local operations and
classical communications (GLOCC) \cite{geof,ordering}. The Gaussian
contangle can be expressed in terms of CMs as
\begin{equation}
\label{gtausig}
G_\tau (\sig) = \inf_{\sig^p \le \sig} E_\tau(\sig^p) \; ,
\end{equation}
where the infimum runs over all pure Gaussian states  with CM
$\sig^p \le \sig$. Let us remark that, if $\sig_{s}$ denotes a mixed
symmetric ($1\times1$)-mode Gaussian state, then the decomposition
of $\sig_s$ in terms of an ensemble of pure Gaussian states is the
optimal one \cite{eofprl}, which means that the Gaussian contangle
coincides with the true contangle. Moreover, the optimal pure-state
CM $\sig_s^p$ minimizing $G_\tau(\sig_s)$ in \eq{gtausig} is
characterized by having $\tilde{\nu}_-(\tilde{\sig}^p_s) =
\tilde{\nu}_-(\tilde{\sig_s})$ \cite{eofprl,geof}. The fact that the
smallest symplectic eigenvalue is the same for both partially
transposed CMs entails for symmetric two-mode Gaussian states that
\begin{equation}
\label{etausym2}
E_\tau(\sig_s) = G_\tau(\sig_s) = [\max\{0,-\ln \tilde
\nu_-(\sig_s)\}]^2 \; .
\end{equation}
Finally, of course $E_\tau = G_\tau$ as well in all {\em pure}
Gaussian states of $1 \times n$ bipartitions.

\subsection{Monogamy inequality for all three-mode Gaussian states}

\label{secmono}

We now provide the detailed proof, first derived, among other results,
in Ref. \cite{contangle}, that all three-mode Gaussian states satisfy
the CKW monogamy inequality (\ref{CKWine}), using the (Gaussian) contangle
to quantify bipartite entanglement. The intermediate steps of the
proof will be then useful for the subsequent computation of the residual
genuine tripartite entanglement, as we will show in Sec. \ref{secresid}.

We start by considering pure three-mode Gaussian states,
whose standard form CM $\sig^p$ is given by \eq{cm3tutta}.
As discussed in Sec. \ref{secpuri}, all the
properties of entanglement in pure three-mode Gaussian states are
completely determined by the three local purities. Reminding that the
mixednesses $a_l \equiv 1/\mu_l$ have to vary constrained
by the triangle inequality (\ref{triangleprim}), in order for
$\sig^p$ to represent a physical state, one has
\begin{equation}
\label{triangle}
|a_j - a_k| + 1 \le a_i \le a_j + a_k - 1 \; .
\end{equation}
For ease of notation let us rename the mode indices so that
$\{i,j,k\} \equiv \{1,2,3\}$.
Without any loss of generality, we can assume $a_1 > 1$.
In fact, if $a_1=1$ the first mode is not correlated with
the other two and all the terms in \ineq{CKWine} are
trivially zero. Moreover, we can restrict the discussion to the
case of both the reduced two-mode states $\sig_{12}$ and $\sig_{13}$
being entangled. In fact, if {\em e.g.}~$\sig_{13}$ denotes a
separable state, then $E_\tau^{1|2} \le E_\tau^{1|(23)}$
because tracing out mode $3$ is a LOCC, and thus the sharing
inequality is automatically satisfied. We
will now prove \ineq{CKWine} in general by using the Gaussian
contangle, as this will immediately imply the inequality for
the true contangle as well. In fact, $G_\tau^{1|(23)}(\sig^p)
= E_\tau^{1|(23)}(\sig^p)$,
but $G_\tau^{1|l}(\sig) \ge E_\tau^{1|l}(\sig),\; l=2,3$.

Let us proceed by keeping $a_1$ fixed.
From \eq{etaupgau}, it follows that the
entanglement between mode $1$ and the remaining modes,
$E_\tau^{1|(23)} = \arcsinh^2\sqrt{a_1^2-1}$, is constant. We must
now prove that the maximum value of the sum of the $1|2$ and $1|3$
bipartite entanglements can never exceed $E_\tau^{1|(23)}$, at fixed
local mixedness $a_1$. Namely,
\begin{equation}
\label{maxQ}
\max_{s,d} Q \, \le \, \arcsinh^2\sqrt{a^{2} - 1} \; ,
\end{equation}
where $a \equiv a_1$ (from now on we drop the subscript ``1''),
and we have defined
\begin{equation}
\label{QQ}
Q \, \equiv \, G_\tau^{1|2}(\sig^p) + G_\tau^{1|3}(\sig^p) \; .
\end{equation}
The maximum in \eq{maxQ} is taken with respect to the ``center of
mass'' and ``relative'' variables $s$ and $d$ that replace the local
mixednesses $a_2$ and $a_3$ according to
\begin{eqnarray}
s &=& \frac{a_2+a_3}2 \; , \label{s1} \\
& & \nonumber \\
d &=& \frac{a_2-a_3}2 \; . \label{d1}
\end{eqnarray}
The two parameters $s$ and $d$ are constrained to vary in the region
\begin{equation}
\label{sdangle}
s \, \ge \, \frac{a+1}{2} \; , \; \; \qquad\abs{d}
\, \le \, \frac{a^2-1}{4s} \; .
\end{equation}
\ineq{sdangle} combines the triangle inequality (\ref{triangle})
with the condition of inseparability for the states of the
reduced bipartitions $(1|2)$ and $(1|3)$ \cite{ordering}.

We recall now, as stated in Sec. \ref{secpuri}, that each
$\sig_{1l}$, $l=2,3$, is a state of partial minimum uncertainty
(GLEMS \cite{extremal}). For this class of states the Gaussian
measures of entanglement, including $G_\tau$, can be computed
explicitely \cite{ordering}, yielding
\begin{widetext}
\begin{equation}
\label{Qglems}
Q = \arcsinh^2 \Big[\sqrt{m^2(a,s,d)-1}\Big]+\arcsinh^2
\Big[\sqrt{m^2(a,s,-d)-1}\Big] \; ,
\end{equation}
where $m = m_-$ if $D \le 0$, and $m = m_+$ otherwise (one has
$m_+=m_-$ for $D=0$). Here:
\begin{eqnarray}
\label{unsacco}
m_- & = & \frac{|k_-|}{(s-d)^2-1} \; , \nonumber \\
& & \nonumber \\
m_+ & = & \frac{\sqrt{2\,\left[2 a^2 (1+2
s^2 + 2 d^2) - (4 s^2 - 1)(4 d^2 - 1) -a^4 -
\sqrt{\delta}\right]}}{4(s-d)}\; , \nonumber \\
& & \nonumber \\
D & = & 2 (s - d) -
\sqrt{2\left[k_-^2 + 2 k_++|k_-| (k_-^2 + 8
k_+)^{1/2}\right]/k_+} \; , \nonumber \\
& & \nonumber \\
k_\pm & = & a^2 \pm (s+d)^2 \; ,
\end{eqnarray}
and the quantity
$$\delta = (a - 2 d - 1) (a - 2 d + 1) (a + 2 d - 1) (a + 2 d +
1) (a - 2 s - 1) (a - 2 s + 1) (a + 2 s - 1) (a + 2 s + 1)$$
\end{widetext}
is the same as in \eq{deltino}. Note (we omitted the
explicit dependence for brevity) that each quantity in \eq{unsacco}
is a function of $(a,s,d)$. Therefore, to evaluate the second term in
\eq{Qglems} each $d$ in \eq{unsacco} must be replaced by $-d$.

Studying the derivative of $m_\mp$ with respect to $s$, it is
analytically proven that, in the whole range of parameters
$\{a,s,d\}$ defined by \ineq{sdangle}, both $m_{-}$ and $m_{+}$
are monotonically decreasing functions of $s$.
The quantity $Q$ is then maximized over $s$ for the limiting value
\begin{equation}
\label{sopt}
s \, = \, s^{\min} \, \equiv \, \frac{a + 1}2 \; .
\end{equation}
This value of $s$ corresponds to three-mode pure Gaussian
states in which the state of the reduced bipartition $2|3$
is always separable, as one should expect because the bipartite
entanglement is maximally concentrated in the states of the
$1|2$ and $1|3$ reduced bipartitions.
With the position \eq{sopt}, the quantity $D$ defined in
\eq{unsacco} can be easily shown to be always negative. Therefore, for
both reduced CMs $\sig_{12}$ and $\sig_{13}$, the Gaussian contangle
is defined in terms of $m_{-}$. The latter, in turn, acquires the
simple form
\begin{equation}
\label{msmin}
m_-(a,s^{\min},d) \, = \, \frac{1 + 3a + 2d}{3 + a - 2 d} \; .
\end{equation}
Consequently, the quantity $Q$ turns out to be an
even and convex function of $d$, and this fact
entails that it is globally maximized at the boundary
\begin{equation}
\label{dopto}
|d| \, = \, d^{\max} \, \equiv \, \frac{a-1}2 \; .
\end{equation}
We finally have that
\begin{eqnarray}
\label{Qmax}
Q^{\max} & \equiv & Q \left[a,s=s^{\min},d=\pm
d^{\max}\right] \nonumber \\
& = & \arcsinh^2 \sqrt{a^2-1} \; ,
\end{eqnarray}
which implies that in this case
the sharing inequality (\ref{CKWine}) is exactly saturated and the
genuine tripartite entanglement is consequently zero. In fact this
case yields states with $a_2=a_1$ and $a_3=1$ (if $d=d^{\max}$), or
$a_3=a_1$ and $a_2=1$ (if $d=-d^{\max}$), {\ie}tensor products of a
two-mode squeezed state and a single-mode uncorrelated vacuum. Being
$Q^{\max}$ from \eq{Qmax} the global maximum of $Q$, \ineq{maxQ}
holds true and the monogamy inequality (\ref{CKWine}) is thus proven
for any pure three-mode Gaussian state, choosing either the Gaussian
contangle $G_\tau$ or the true contangle $E_\tau$ as measures of
bipartite entanglement \cite{contangle}.

The proof immediately extends to all mixed three-mode Gaussian
states $\sig$, but only if the bipartite entanglement is measured by
$G_\tau(\sig)$ \cite{Footnote}. Let $\{\pi(d\sig^p_{m}),
\sig^{p}_{m}\}$ be the ensemble of pure Gaussian states minimizing
the Gaussian convex roof in Eq.~(\ref{GaCoRo}); then, we have
\begin{eqnarray}
\label{convext}
G_\tau^{i|(jk)}(\sig) &=& \int \pi (d\sig^p_{m})
G_\tau^{i|(jk)}(\sig^p_{m}) \nonumber \\
& \ge & \int \pi (d\sig^p_{m})
[G_\tau^{i|j}(\sig^p_{m}) + G_\tau^{i|k}(\sig^p_{m})] \\
&\ge& G_\tau^{i|j}(\sig) + G_\tau^{i|k}(\sig) \; , \nonumber
\end{eqnarray}
where we exploited the fact that the Gaussian contangle is convex by
construction. This concludes the proof of the CKW monogamy
inequality~(\ref{CKWine}) for all three-mode Gaussian states.

We close this subsection by discussing whether the CKW monogamy
inequality can be generalized to all Gaussian states of systems with
an arbitrary number $n+1$ of modes. Namely, we want to prove that
\begin{equation}
\label{monoN} E^{i|(j_1 , \ldots , j_n)} - \sum_{l=1}^{n} E^{i|j_l}
\geq 0 \; .
\end{equation}
Establishing this result in general is a highly nontrivial task, but
it can be readily proven for all {\it symmetric} multimode Gaussian
states \cite{contangle}. In a fully symmetric $n+1$-mode Gaussian
state all the local purities are degenerate and reduce to a single
parameter $a_{loc}$:
\begin{equation}
a_{i} \, = \, a_{j_{1}} \, = \, a_{j_{2}} \, = \, \dots \, = \,
a_{j_{n}} \, \equiv \, a_{loc} \; . \label{symmetricmultimode}
\end{equation}
As in the three-mode case, due to the convexity of $G_\tau$, it will
suffice to prove \eq{monoN} for pure states, for which the Gaussian
contangle coincides with the true contangle in every bipartition.
For any $n$ and for $a_{loc}> 1$ (for $a_{loc}=1$ we have a product
state), one has that
\begin{equation}
E_\tau^{i|(j_1 , \ldots , j_n)} = \ln^2(a_{loc}-\sqrt{a_{loc}^2-1})
\end{equation}
is independent of $n$, while the total two--mode contangle
\begin{eqnarray}
n E_\tau^{i|j_l} & = & \frac{n}{4} \ln^2
\bigg\{ \Big[ a_{loc}^2 (n+1) - 1  \nonumber \\
& - &  \sqrt{(a_{loc}^2-1)(a_{loc}^2(n+1)^2-(n-1)^2)}
\; \; \Big] /n \bigg\} \nonumber \\
& &
\end{eqnarray}
is a monotonically decreasing function of the integer $n$ at fixed
$a_{loc}$. Because the sharing inequality trivially holds for $n=1$,
it is inductively proven for any $n$. This result, together with
extensive numerical evidence obtained for randomly generated
non-symmetric $4$--mode Gaussian states \cite{contangle}, strongly
supports the conjecture that the CKW monogamy inequality holds true
for {\em all} multimode Gaussian state, using the (Gaussian)
contangle as a measure of bipartite entanglement. However, at
present, a fully analytical proof of this conjecture is still
lacking.

\subsection{Residual contangle, genuine tripartite entanglement,
            and monotonicity}

\label{secresid}

\begin{figure}[t!]
\includegraphics[width=8.5cm]{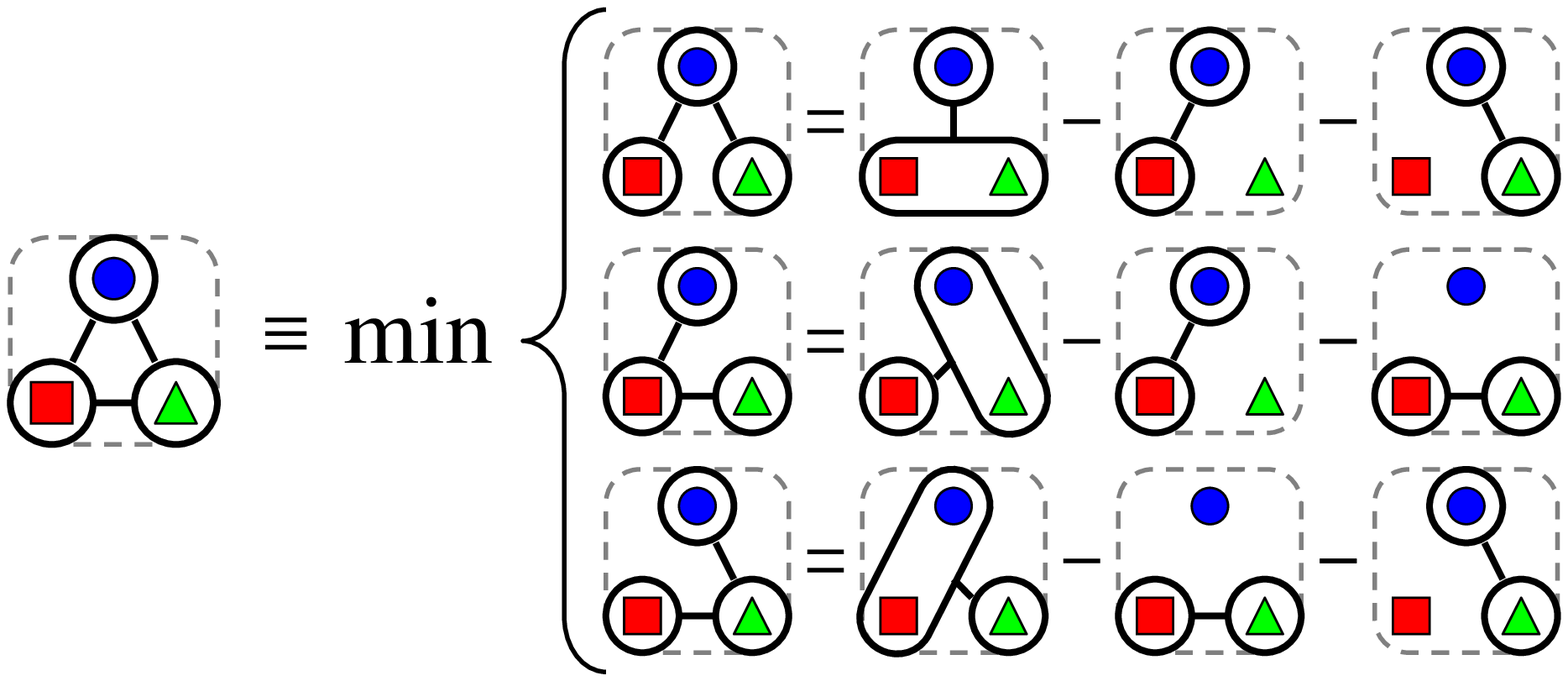}
\centering{\caption{(color online) Pictorial representation of
\eq{gtaures}, defining the residual Gaussian contangle
$G_\tau^{res}$ of generic (nonsymmetric) three-mode Gaussian states.
$G_\tau^{res}$ quantifies the genuine tripartite entanglement shared
among mode $1$ ({\footnotesize{$\color[rgb]{0,0,1}{\Circle[f]}$}}),
mode $2$ ({\tiny{$\color[rgb]{1,0,0}{\blacksquare}$}}), and mode $3$
({\scriptsize{$\color[rgb]{0,1,0}{\blacktriangle}$}}). The optimal
decomposition that realizes the minimum in \eq{gtaures} is always
the one for which the CM of the reduced state of the reference mode
has the smallest determinant.} \label{figomini}}
\end{figure}

The sharing constraint leads naturally to the definition of the {\em
residual contangle} as a quantifier of genuine tripartite
entanglement ({\it arravogliament}) in three-mode Gaussian states,
much in the same way as in systems of three qubits \cite{CKW}.
However, at variance with the three-qubit case, here the residual
contangle is partition-dependent according to the choice of the
reference mode, with the exception of the fully symmetric states. A
{\em bona fide} quantification of tripartite entanglement is then
provided by the {\em minimum} residual contangle \cite{contangle}
\begin{equation}
\label{etaumin}
E_\tau^{i|j|k}\equiv\min_{(i,j,k)} \left[
E_\tau^{i|(jk)}-E_\tau^{i|j}-E_\tau^{i|k}\right] \; ,
\end{equation}
where the symbol $(i,j,k)$ denotes all the permutations of the three
mode indexes. This definition ensures that $E_\tau^{i|j|k}$ is
invariant under all permutations of the modes and is thus a genuine
three-way property of any three-mode Gaussian state. We can adopt an
analogous definition for the minimum residual Gaussian contangle
$G_\tau^{res}$ (see Fig. \ref{figomini} for a pictorial
representation):
\begin{equation}
\label{gtaures}
G_\tau^{res} \equiv G_\tau^{i|j|k}\equiv\min_{(i,j,k)} \left[
G_\tau^{i|(jk)}-G_\tau^{i|j}-G_\tau^{i|k}\right] \; .
\end{equation}
One can verify that
\begin{equation}
\label{refsat}
(G_\tau^{i|(jk)} \, - \, G_\tau^{i|k}) \, - \,
(G_\tau^{j|(ik)} \, - \, G_\tau^{j|k}) \, \ge \, 0
\end{equation}
if and only if $a_i \ge a_j$, and therefore the absolute minimum
in \eq{etaumin} is attained by the decomposition realized with
respect to the reference mode $l$ of smallest local mixedness $a_l$,
i.e. for the single-mode reduced state with CM of smallest determinant.

The residual (Gaussian) contangle must be nonincreasing under
(Gaussian) LOCC in order to be a proper measure of tripartite
entanglement. The monotonicity of the residual tangle was proven for
three-qubit pure states in Ref. \cite{wstates}. In the CV setting,
it has been shown in Ref. \cite{contangle} that for pure three-mode
Gaussian states the residual Gaussian contangle \eq{gtaures} is an
entanglement monotone under tripartite GLOCC, and that it is
nonincreasing even under probabilistic operations, which is a
stronger property than being only monotone on average. Therefore the
Gaussian contangle $G_\tau^{res}$ defines (to the best of our
knowledge) {\em the first} measure, proper and computable, of
genuine multipartite (specifically, tripartite) entanglement in
 Gaussian states of CV systems. It is worth noting that the
{\em minimum} in \eq{gtaures}, that at first sight might appear a
redundant requirement, is physically meaningful and mathematically
necessary. In fact, if one chooses to fix a reference partition, or
to take {\eg}the maximum (and not the minimum) over all possible
mode permutations in \eq{gtaures}, the resulting ``measure'' is not
monotone under GLOCC and thus is definitely {\em not} a measure of
tripartite entanglement.

We now work out in detail an explicit application, by describing
the complete procedure to determine the genuine tripartite
entanglement in a {\em pure} three-mode Gaussian state $\sig^p$.

\begin{description}

\item[{\rm (i)} Determine the local purities.] The
state is globally pure ($\det\sig^p = 1$); therefore,
the only quantities needed for the computation of
the tripartite entanglement are the three local
mixednesses $a_l$, defined by \eq{al}, of the
single-mode reduced states $\sig_l,\,l=1,2,3$
(see \eq{subma}). Notice that the global CM
$\sig^p$ needs not to be in the standard form (\ref{cm3tutta}), as
the single-mode determinants are local symplectic invariants
\cite{serafozzi}. From an experimental point of view, the parameters
$a_l$ can be extracted from the CM using the homodyne tomographic
reconstruction of the state \cite{homotomo}; or they can be
directly measured with the aid of single photon detectors
\cite{fiuracerf,wenger}.

\item[{\rm (ii)} Find the minimum.] From \eq{refsat}, the minimum in the definition
(\ref{gtaures}) of the residual Gaussian contangle $G_\tau^{res}$ is attained
in the partition where the bipartite entanglements are decomposed
choosing as reference mode $l$ the one in the single-mode reduced state
of smallest local mixedness $a_l \equiv a_{min}$.

\item[{\rm (iii)} Check range and compute.] Given the mode
with smallest local mixedness $a_{min}$ (say, for instance,
mode $1$) and the parameters $s$ and $d$ defined in
Eqs.~(\ref{s1},\ref{d1}), if $a_{min}=1$
then mode $1$ is uncorrelated from the others:
$G_\tau^{res}=0$. If, instead, $a_{min}>1$ then
\begin{equation}
\label{gtaurespur}
G_\tau^{res} (\sig^p) =  \arcsinh^2\!\Big[\sqrt{a_{min}^2-1}\Big]
- Q(a_{min},s,d) \; ,
\end{equation}
with $Q \equiv G_\tau^{1|2} + G_\tau^{1|3}$ defined by
Eqs.~(\ref{Qglems},\ref{unsacco}).
Note that if $d<-(a_{min}^2-1)/4s$
then $G_\tau^{1|2}=0$. Instead, if $d>(a_{min}^2-1)/4s$ then
$G_\tau^{1|3}=0$. Otherwise, all terms
in $G_\tau^{res}$ \eq{gtaures} are nonvanishing.
\end{description}

\begin{figure}[t!]
\centering{\includegraphics[width=8.5cm]{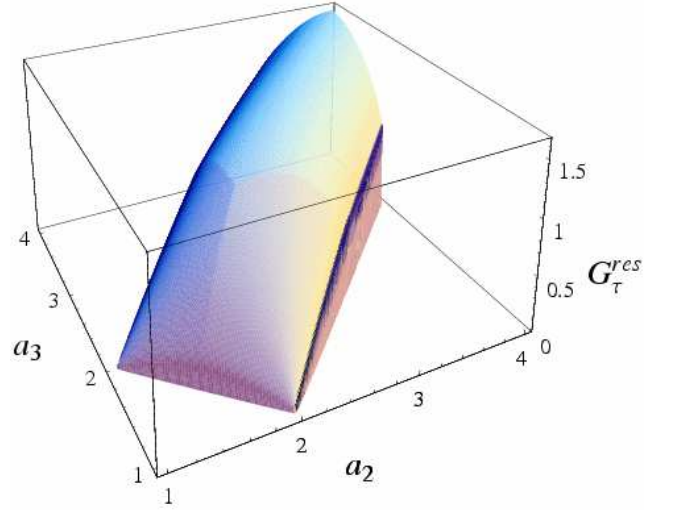}%
\caption{\label{figsupposta}(color online) Three-dimensional plot of
the residual Gaussian contangle $G_\tau^{res}(\sig^p)$ in pure
three-mode Gaussian states $\sig^p$, determined by the three local
mixedness $a_l$, $l=1,2,3$. One of the local mixedness is kept fixed
($a_1=2$). The remaining ones vary constrained by the triangle
inequality (\ref{triangle}), as depicted in Fig. \ref{figangle}(b).
The explicit expression of $G_\tau^{res}$ is given by \eq{glems}.
See text for further details.}}
\end{figure}
The residual Gaussian contangle \eq{gtaures} in generic pure
three-mode Gaussian states is plotted in Fig. \ref{figsupposta} as a
function of $a_2$ and $a_3$, at constant $a_1=2$. For fixed $a_1$,
it is interesting to notice that $G_\tau^{res}$ is maximal for
$a_2=a_3$, {\ie}for bisymmetric states. Notice also how the residual
Gaussian contangle of these bisymmetric pure states has a cusp for
$a_1=a_2=a_3$. In fact, from \eq{refsat}, for $a_2=a_3 < a_1$ the
minimum in \eq{gtaures} is attained decomposing with respect to one
of the two modes $2$ or $3$ (the result is the same by symmetry),
while for $a_2=a_3 > a_1$ mode $1$ becomes the reference mode.

For generic {\em mixed} three-mode  Gaussian states, a quite
cumbersome analytical expression for the $1|2$ and $1|3$ Gaussian
contangles may be written \cite{geof,ordering}, involving the roots
of a fourth order polynomial, but the optimization appearing in the
computation of the  $1|(23)$ bipartite Gaussian contangle (see
\eq{gtausig}) has to be solved only numerically. However, exploiting
techniques like the unitary localization \cite{unitarily} described
in Sec. \ref{secsym}, and results like that of \eq{etausym2}, closed
expressions for the residual Gaussian contangle can be found as well
in relevant classes of mixed three-mode Gaussian states endowed with
some symmetry constraints. Interesting examples of these states and
the investigation of their physical properties will be discussed in
Sec. \ref{secstructex}.

As an additional remark, let us recall that, although the
entanglement of Gaussian states is always distillable with respect
to $1\times N$ bipartitions \cite{werwolf}, they can exhibit bound
entanglement in $1 \times 1 \times  1$ tripartitions \cite{kraus}.
In this case, the residual contangle cannot detect
 tripartite PPT entangled states. For example, the
residual contangle in three-mode biseparable Gaussian states (class
$4$ of Ref. \cite{kraus}) is always zero, because those bound
entangled states are separable with respect to all $(1\times2)$-mode
bipartitions. In this sense we can correctly regard the residual
contangle as an estimator of {\em distillable} tripartite
entanglement in fully inseparable three-mode Gaussian states.
However, we remind that this entanglement can be distilled only
resorting to non-Gaussian LOCC \cite{browne}, since distilling
Gaussian states with Gaussian operations is impossible
\cite{nogo1,nogo2,nogo3}.

\section{Sharing structure of tripartite entanglement}\label{secstructex}

We are now in the position to analyze the sharing structure of CV
entanglement in three-mode Gaussian states
by taking the residual Gaussian contangle as a measure
of tripartite entanglement, in analogy with the study done for three
qubits \cite{wstates} using the residual tangle \cite{CKW}.

The first task we face is that of identifying the three-mode
analogues of the two inequivalent classes of fully inseparable
three-qubit states, the GHZ state \cite{ghzs}
\begin{equation}
\label{qghz}
\ket{\psi_{\rm GHZ}} \, = \, \frac{1}{\sqrt2} \left(\ket{000} +
\ket{111}\right) \; ,
\end{equation}
and the $W$ state \cite{wstates}
\begin{equation}
\label{qghz2}
\ket{\psi_{W}} \, = \, \frac{1}{\sqrt3} \left(\ket{001} + \ket{010} +
\ket{100}\right) \; .
\end{equation}
These states are both pure and fully symmetric, {\ie}invariant under
the exchange of any two qubits. On the one hand, the GHZ state
possesses maximal tripartite entanglement, quantified by the
residual tangle \cite{CKW,wstates}, with zero couplewise
entanglement in any reduced state of two qubits reductions.
Therefore its entanglement is very fragile against the loss of one
or more subsystems. On the other hand, the $W$ state contains the
maximal two-party entanglement in any reduced state of two qubits
\cite{wstates} and is thus maximally robust against decoherence,
while its tripartite residual tangle vanishes.

\subsection{CV GHZ/$W$ states}\label{secghzw}

To define the CV counterparts of the three-qubit states
$\ket{\psi_{\rm GHZ}}$ and $\ket{\psi_{W}}$, one must start from the
fully symmetric three-mode CM $\sig_s$ of \eq{fscm}. Surprisingly
enough, in symmetric three-mode Gaussian states, if one aims at
maximizing, at given single-mode mixedness $a_{loc} \equiv a$,
either the bipartite entanglement $G_\tau^{i|j}$ in any two-mode
reduced state ({\it i.e.}~aiming at the CV $W$-like state), or the
genuine tripartite entanglement $G_\tau^{res}$ ({\it i.e.}~aiming at
the CV GHZ-like state), one finds the same, unique family of pure
symmetric three-mode squeezed states $\sig^{p}_{s}$. These states,
previously known as CV ``GHZ-type'' states
\cite{network,bravchap,vanlokfuru}, can be indeed defined for
generic $n$-mode systems. They constitute an ideal test-ground for
the study of the scaling of multimode CV entanglement with the
number of modes. This analysis can be carried out via nested
applications of the procedure of unitary localization
\cite{adescaling,unitarily}, reviewed in Sec. \ref{secsym}. For
systems of three modes, they are described by a CM $\sig_s^p$ of the
form \eq{fscm}, with $\gr\alpha=a \id_2$, $\gr\varepsilon={\rm
diag}\{e^+,\,e^-\}$ and \cite{adescaling}
\begin{equation}
\label{epmfulsym}
e^\pm = \frac{a^2-1 \pm \sqrt{\left(a^2 - 1\right) \left(9 a^2 -
1\right)}}{4a} \; ,
\end{equation}
ensuring the global purity of the state. For self-explaining
reasons, we choose to name these states ``CV GHZ/$W$ states''
\cite{contangle}, and denote their CM by $\sig_{s}^{GHZ/W}$.
In the limit of infinite squeezing ($a \rightarrow
\infty$), the CV GHZ/$W$ state approaches the proper (unnormalizable)
continuous-variable GHZ state $\int dx \ket{x,x,x}$, a
simultaneous eigenstate of total momentum
$\hat{p}_1+\hat{p}_2+\hat{p}_3$ and of all relative positions
$\hat{x}_i - \hat{x}_j$ ($i,j=1,2,3$), with zero eigenvalues
\cite{cvghz}.

The residual Gaussian contangle of GHZ/$W$ states
of finite squeezing takes the simple form
\cite{contangle}
\begin{equation}
\label{gresghzw}
\begin{split}
G_\tau^{res}(\sig_{s}^{GHZ/W})&= \arcsinh^2\!\left[\sqrt{a^2 -1}\right] \\
&- \frac{1}{2} \ln^2\!\left[\frac{3 a^2 - 1 -\sqrt{9 a^4 - 10 a^2 +
1}}{2}\right]\, .
\end{split}
\end{equation}
It is straightforward to see that $G_\tau^{res}(\sig_s^p)$
is nonvanishing as soon as $a>1$. Therefore, the GHZ/$W$
states belong to the class of fully inseparable three-mode
states \cite{kraus,adescaling,network,vloock02,vanlokfuru}.
We finally recall that in a GHZ/$W$ state the residual Gaussian
contangle $G_\tau^{res}$ \eq{gtaures} coincides with the true
residual contangle $E_\tau^{res}$ \eq{etaumin}. This property
clearly holds because the Gaussian pure-state decomposition
is the optimal one in every bipartition, due to the fact that
the global three-mode state is pure and the reduced two-mode states
are symmetric.

\subsection{$T$ states}\label{sectstates}

The peculiar nature of entanglement sharing in  CV GHZ/$W$ states is
further confirmed by the following observation. If one requires
maximization of the $1 \times 2$ bipartite Gaussian contangle
$G_\tau^{i|(jk)}$ under the constraint of separability of all the
reduced two-mode states, one finds a class of symmetric mixed states
characterized by being three-mode Gaussian states of partial minimum
uncertainty. They are in fact characterized by having their smallest
symplectic eigenvalue equal to $1$, and represent thus the
three-mode generalization of two-mode symmetric GLEMS
\cite{prl,extremal,ordering}.

We will name these states {\em $T$ states}, with $T$ standing for {\em
tripartite} entanglement only. They are described by a CM $\sig_s^T$
of the form \eq{fscm}, with $\gr\alpha=a \id_2$,
$\gr\varepsilon={\rm diag}\{e^+,\,e^-\}$ and
\begin{eqnarray}
\label{epmtstat}
e^+  & = & \frac{a^2 -5 + \sqrt{9  a^2 \left(a^2 - 2\right) + 25}}{4a}
\; , \nonumber \\
e^- & = & \frac{5-9a^2 + \sqrt{9  a^2 \left(a^2 -
2\right) + 25}}{12 a} \; .
\end{eqnarray}
The $T$ states, like the GHZ/$W$ states, are determined only by the
local mixedness $a$, are fully separable for $a=1$, and fully
inseparable for $a>1$. The residual Gaussian contangle \eq{gtaures}
can be analytically computed for these mixed states as a function of
$a$. First of all one notices that, due to the complete symmetry of
the state, each mode can be chosen indifferently to be the reference
one in \eq{gtaures}. Being the $1 \times 1$ entanglements all zero
by construction, $G_\tau^{res} = G_\tau^{i|(jk)}$. The $1 \times 2$
bipartite Gaussian contangle can be in turn obtained exploiting the
unitary localization procedure (see Sec. \ref{secsym}). Let us
choose mode $1$ as the reference mode and combine modes $2$ and $3$
at a 50:50 beam splitter, a local unitary operation with respect to
the bipartition $1|(23)$ that defines the transformed modes $2'$ and
$3'$. The CM $\sig_s^{T'}$ of the state of modes $1$, $2'$, and $3'$
is then written in the following block form:
\begin{equation}
\label{sigtprim}
\sig_s^{T'} = \left(\begin{array}{ccc}
\sig_{1} & \eps_{12'} & {\bf 0} \\
\eps_{12'}^{\sf T} & \sig_{2'} & {\bf 0} \\
{\bf 0} & {\bf 0} & \sig_{3'} \\
\end{array}\right) \; ,
\end{equation}
where mode $3'$ is now disentangled from the others. Thus
\begin{equation}
\label{tlocato}
G_\tau^{1|(23)}(\sig_s^T) = G_\tau^{1|2'}(\sig_s^{T'}) \; .
\end{equation}
Moreover, the reduced CM $\sig_{12'}$ of modes $1$ and $2'$ defines
a nonsymmetric GLEM \cite{prl,extremal} with
\begin{eqnarray*}
\det\sig_1 &=&a^2\,, \\
\det\sig_2 &=& \frac{1}{6} \left(3 a^2 +
\sqrt{9 \left(a^2 - 2\right) a^2 + 25} - 1\right)\,, \\
\det\sig_{12'} &=& \frac{1}{2} \left(3 a^2
- \sqrt{9 \left(a^2 - 2\right) a^2 + 25} + 3\right)\,,
\end{eqnarray*}
and it has been shown that the Gaussian contangle is
computable in two-mode GLEMS \cite{ordering}.
After some algebra, one finds the complete
expression of $G_\tau^{res}$ for $T$ states:
\begin{eqnarray}
\label{grest}
G_\tau^{res}(\sig_s^T) &=& \arcsinh^2 \Bigg\{ \bigg[25 R -9 a^4 + 3
R a^2
+ 6 a^2  -109 \nonumber \\
&-& \Big(81 a^8 - 432 a^6 + 954 a^4 -
1704 a^2 +2125 \nonumber \\
&-& \left(3 a^2 - 11\right) \left(3 a^2 -
7\right) \left(3 a^2 + 5\right)
R\Big)^{\frac12}\sqrt{2}\bigg]^{\frac12} \nonumber \\
&\times& \left[{18 \left(3 a^2 - R +
3\right)}\right]^{-\frac12}\Bigg\} \; ,
\end{eqnarray}
with $R \equiv \sqrt{9 a^2 (a^2 - 2) + 25}$.

\begin{figure}[t!]
\centering{
\includegraphics[width=8.5cm]{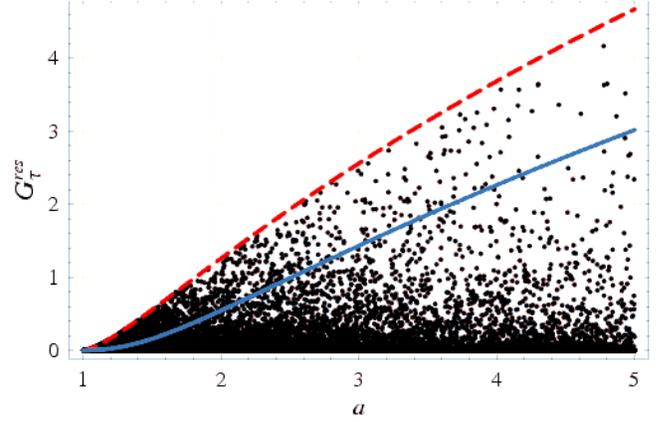}
\caption{(color online) Plot, as a function of the single-mode
mixedness $a$, of the tripartite residual Gaussian contangle
$G_\tau^{res}$ \eq{gresghzw} in the CV GHZ/$W$ states (dashed line);
in the $T$ states \eq{grest} (solid line); and in 50000 randomly
generated mixed symmetric three-mode Gaussian states of the form
\eq{fscm} (dots). The GHZ/$W$ states, that maximize any bipartite
entanglement, also achieve maximal genuine tripartite quantum
correlations, showing that CV entanglement distributes in a
promiscuous way in symmetric Gaussian states. Notice also how all
random mixed states have a nonnegative residual Gaussian contangle.
This confirms the results presented in Ref.~\cite{contangle}, and
discussed in detail and extended in Sec.~\ref{secmono}, on the
strict validity of the CKW monogamy inequality for CV entanglement
in three-mode Gaussian states.} \label{figatua}}
\end{figure}

What is remarkable about $T$ states is that their tripartite
Gaussian contangle \eq{grest} is strictly smaller than the one of
the GHZ/$W$ states \eq{gresghzw} for any fixed value of the local
mixedness $a$, that is, for any fixed value of the only parameter
(operationally related to the squeezing of each single mode) that
completely determines the CMs of both families of states up to local
unitary operations. This hyerardical behavior of the residual
contangle in the two classes of states is illustrated in Fig.
\ref{figatua}. Notice that this result cannot be an artifact caused
by restricting to pure Gaussian decompositions only in the
definition \eq{gtaures} of the residual Gaussian contangle. In fact,
for $T$ states the relation $G_\tau^{res}(\sig_s^T) \ge
E_\tau^{res}(\sig_s^T)$ holds due to the symmetry of the reduced
two-mode states, and to the fact that the unitarily transformed
state of modes $1$ and $2'$ is mixed and nonsymmetric. The crucial
consequences of this result for the structure of the entanglement
trade-off in Gaussian states will be discussed further in the next
subsection.

\subsection{Promiscuous continuous-variable entanglement
sharing} \label{secpromis}

The above results, pictorially illustrated in Fig. \ref{figatua},
lead to the conclusion that in symmetric three-mode Gaussian states,
when there is no bipartite entanglement in the two-mode reduced
states (like in $T$ states) the genuine tripartite entanglement is
not enhanced, but frustrated. More than that, if there are maximal
quantum correlations in a three-party relation, like in GHZ/$W$
states, then the two-mode reduced states of any pair of modes are
maximally entangled mixed states.

These findings, unveiling a major difference between
discrete-variable (mainly qubits) and continuous-variable systems,
establish the {\em promiscuous} nature of CV entanglement sharing in
symmetric Gaussian states \cite{contangle,sharing}. Being associated
with degrees of freedom with continuous spectra, states of CV
systems need not saturate the CKW inequality to achieve maximum
couplewise correlations. In fact, without violating the monogamy
constraint \ineq{CKWine}, pure symmetric three-mode Gaussian states
are maximally three-way entangled and, at the same time, maximally
robust against the loss of one of the modes. This preselects GHZ/$W$
states also as optimal candidates for carrying quantum information
through a lossy channel, being, for their intrinsic entanglement
structure, less sensitive to decoherence effects, as we will show
in Sec. \ref{decoherence}.

As an additional remark, let us mention that, quite naturally, not
all three-mode Gaussian states (in particular nonsymmetric states)
are expected to exhibit a promiscuous entanglement sharing. Further
investigations to clarify the sharing structure of generic Gaussian
states of CV systems, and the origin of the promiscuity, are
currently under way \cite{3mjpart1}. As an anticipation, we can
mention that promiscuity tends to survive even in the presence of
mixedness of the state, but is destroyed by the loss of complete
symmetry. The powerful consequences of the entanglement
properties of GHZ/$W$ states for experimental implementations of CV
quantum-information protocols are currently under investigation
\cite{3mjpart2}.

\section{Decoherence of three-mode states and decay of tripartite entanglement}
\label{decoherence}

Remarkably, Gaussian states allow for a straightforward, analytical
treatment of decoherence, accounting for the most common situations
encountered in the experimental practice (like fibre propagations or
cavity decays) and even for more general, `exotic' settings (like
``squeezed'' or common reservoirs) \cite{serafozzijob05}. This
agreeable feature, together with the possibility -- extensively
exploited in this paper -- of exactly computing several interesting
benchmarks for such states, make Gaussian states a useful
theoretical reference for investigating the effect of decoherence on
the information and correlation content of quantum states. Let us
mention that the dissipative evolution of three-mode states has been
considered in Ref.~\cite{paris05}, addressing SU(2,1) coherent
states and focusing essentially on separability thresholds and
telecloning efficiencies. In this section, we will explicitly show
how the decoherence of three-mode Gaussian states may be exactly
studied for any finite temperature, focusing on the evolution of the
residual contangle as a measure of tripartite correlations. The
results here obtained will be recovered in future work
\cite{3mjpart2}, and applied to the study of the effect of
decoherence on multiparty protocols of CV quantum communication with
the classes of states we are addressing, thus completing the present
analysis by investigating its precise operational consequences.
Concerning the general theory of open quantum dynamics, it is impossible
here to give a detailed account of all the aspects of the standard
theoretical frameworks. For an excellent critical review, focusing on
the standard treatment of open quantum systems in relation to quantum
entanglement see Ref.~\cite{Benatti}. In this ample review the authors
discuss the importance of notions such as complete positivity, a physically
motivated algebraic constraint on the quantum dynamics, in relation to
quantum entanglement, and analyze the entanglement power of heat baths
versus their decohering properties.

For continuous-variable systems,
in the most customary and relevant instances the bath interacting with a
set of $n$ modes can be modeled by $n$ independent continua of
oscillators, coupled to the bath through a quadratic Hamiltonian
$H_{int}$ in rotating wave approximation, reading
\be
H_{int}=\sum_{i=1}^{n} \int
v_i(\omega)[a_i^{\dag}b_i(\omega)+a_ib_i^{\dag}(\omega)] \,{\rm
d}\omega \, ,
\label{coupling}
\ee
where $b_i(\omega)$ stands for
the annihilation operator of the $i$th continuum's mode labeled by
the frequency $\omega$, whereas $v_i(\omega)$ represents the
coupling of such a mode to the mode $i$ of the system (assumed, for
simplicity, to be real). The state of the bath is assumed to be
stationary. Under the Born-Markov approximation \cite{bornmarkov},
the Hamiltonian $H_{int}$ leads, upon partial tracing over the bath,
to the following master equation for the $n$ modes of the system (in
interaction picture) \cite{carmichael}
\be
\dot\varrho\;  = \;
\sum_{i=1}^{n} \frac{\gamma_i}{2}\Big(N_i \: L[a_i^{\dag}]\varrho
+(N_i+1)\:L[a_i]\varrho \Big) \, ,
\label{rhoev}
\ee
where the dot stands for time--derivative, the Lindblad superoperators are defined
as\index{Lindblad superoperators} $L[\hat{o}]\varrho \equiv  2
\hat{o}\varrho \hat{o}^{\dag} - \hat{o}^{\dag} \hat{o}\varrho
-\varrho \hat{o}^{\dag} \hat{o}$, the couplings are $\gamma_i=2\pi
v_i^{2}(\omega_i)$, whereas the coefficients $N_i$ are defined in
terms of the correlation functions $\langle
b_i^{\dag}(\omega_i)b_i(\omega_i) \rangle = N_i$, where averages are
computed over the state of the bath and $\omega_i$ is the frequency
of mode $i$. Notice that $N_i$ is the number of thermal photons
present in the reservoir associated to mode $i$, related to the
temperature $T_i$ of the reservoir by the Bose statistics at null
chemical potential: \be N_i =
\frac{1}{{\exp}({\frac{\omega_i\hbar}{kT_{i}}})-1} \; . \label{bose}
\ee In the derivation, we have also assumed $\langle
b_i(\omega_i)b_i(\omega_i) \rangle = 0$, holding for a bath at
thermal equilibrium. We will henceforth refer to a ``homogeneous''
bath in the case $N_{i}=N$ and $\gamma_i=\gamma$ for all $i$.

Now, the master equation (\ref{rhoev}) admits a simple and
physically transparent representation as a diffusion equation for
the time-dependent characteristic function of the system
$\chi(\xi,t)$ \cite{carmichael} \begin{widetext}\be
\dot{\chi}(\xi,t)=-\sum_{i=1}^{n}\frac{\gamma_i}{2}\Bigg[ (x_i\;
p_i) {\partial{x_i}\choose \partial{p_i}} + (x_i\; p_i)
\gr\omega^{\sf T}\gr{\sigma}_{i\infty}\gr\omega {x_i\choose p_i}
\Bigg] \chi(\xi,t) \, , \label{fokpla} \ee \end{widetext} where
$\xi\equiv(x_1,p_1,\ldots,x_n,p_n)$ is a phase-space vector,
$\sig_{i\infty}=\,{\rm diag}\,(2N_i+1,2N_i+1)$ and ${\gr \omega}$ is the
$2\times 2$ symplectic form [defined in \eq{symform}].
The right hand side of the previous equation contains a
deterministic drift term, which has the effect of damping the first
moments to zero on a time scale of $\gamma/2$ and a diffusion term
with diffusion matrix
$\sig_{\infty}\equiv\oplus_{i=1}^{n}\sig_{i\infty}$. The essential
point here is that \eq{fokpla} preserves the Gaussian character of
the initial state, as can be straightforwardly checked for any
initial CM $\sig_0$ by inserting the Gaussian characteristic
function $\chi(\xi,t)$ [see \eq{gauss}]
$$
\chi(\xi,t) = \,{\rm e}^{-\frac12 \xi^{\sf T}\Omega^{\sf T}\sig(t)\Omega\xi
+i X^{\sf T}\Gamma_t\Omega \xi}\;
$$
(where $X$ are generic initial first moments,
$\sig(t)\equiv\Gamma_t^2\sig_0+(\id-\Gamma_t^2)\sig_{\infty}$ and
$\Gamma_{t}\equiv\oplus_{i}{\rm e}^{-\gamma_i t/2}\id_2$) into the
equation and verifying that it is indeed a solution. Notice that,
for a homogeneous bath, the diagonal matrices $\Gamma_t$ and
$\sig_{\infty}$ (providing a full characterisation of the bath)
are both proportional to the identity.
In order to keep track of the decay of
correlations of Gaussian states, we are interested in the evolution
of the initial CM $\sig_{0}$ under the action of the bath which,
recalling our previous Gaussian solution, is just described by
\be
\sig (t) = \Gamma_{t}^2\sig_0+(\id-\Gamma_t^2)\sig_{\infty} \;
\label{cmevo} \ee
This simple equation describes the dissipative
evolution of the CM of any initial state under the action of a
thermal environment and, at zero temperature, under the action of
``pure losses'' (recovered in the instance $N_i=0$ for
$i=1,\ldots,n$). It yields a basic, significant  example of
`Gaussian channel', {\em i.e.}~of a map mapping Gaussian states into
Gaussian states under generally non unitary evolutions. Exploiting
\eq{cmevo} and our previous findings, we can now study the exact
evolution of the tripartite entanglement of Gaussian states under
the decoherent action of losses and thermal noise. For simplicity,
we will mainly consider homogeneous baths.

As a first general remark let us notice that, in the case of a zero
temperature bath ($N=0$), in which decoherence is entirely due to
losses, the bipartite entanglement between any different partitions
decays in time but persists for an infinite time. This is a general
property of Gaussian entanglement \cite{serafozzijob05} under any
many mode bipartition. The same fact is also true for the genuine
tripartite entanglement, quantified by the residual contangle. If
$N\neq0$, a finite time does exist for which tripartite quantum
correlations disappear. In general, the two-mode entanglement
between any given mode and any other of the remaining two modes
vanishes before than the three-mode bipartite entanglement between
such a mode and the other two [not surprisingly, as the former
quantity is, at the beginning, bounded by the latter because of the
CKW monogamy inequality (\ref{CKWine})].

\begin{figure}[t!]
\centering{
\includegraphics[width=8.5cm]{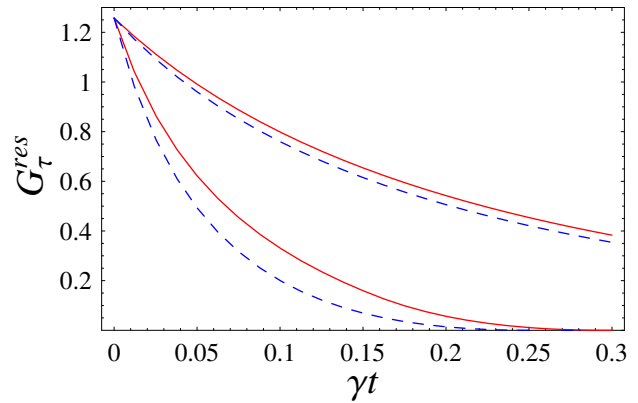}
\caption{(color online) Evolution of the residual Gaussian contangle
$G^{res}_{\tau}$ for GHZ/$W$ states with local mixedness $a=2$
(solid curves) and $T$ states with local mixedness $a=2.8014$
(dashed curves). Such states have equal initial residual contangle.
The uppermost curves refer to a homogeneous bath with $N=0$ (pure
losses), while the lowermost curves refer to a homogeneous bath with
$N=1$. As apparent, thermal photons are responsible for the
vanishing of entanglement at finite times.} \label{decofig1}}
\end{figure}

The main issue addressed in this analysis has consisted in
inspecting the robustness of different forms of genuine tripartite
entanglement, previously introduced in the paper. Notice that an
analogous question has been addressed in the qubit scenario, by
comparing the action of decoherence on the residual tangle of the
inequivalent sets of GHZ and $W$ states: $W$ states, which are by
definition more robust under subsystem erasure, proved more robust
under decoherence as well \cite{carvalho04}. In our instance, the
symmetric GHZ/$W$ states constitute a promising candidate for the
role of most robust Gaussian tripartite entangled states, as
somehow expected. Evidence supporting this conjecture is shown in
Fig.~\ref{decofig1}, where the evolution in different baths of the
tripartite entanglement of GHZ/$W$ states is compared to that of
symmetric $T$ states (at same initial entanglement). No fully
symmetric states with tripartite entanglement more robust than
GHZ/$W$ states were found by further numerical inspection.
Quite remarkably, the promiscuous sharing of quantum correlations,
proper to GHZ/W states, appears to better preserve genuine multipartite
entanglement against the action of decoherence.

Notice also that, for a homogeneous bath and for all fully symmetric
and bisymmetric three-mode states, the decoherence of the global
{\em bipartite} entanglement of the state is the same as that of the
corresponding equivalent two-mode states (obtained through unitary
localization). Indeed, for any bisymmetric state which can be
localized by an orthogonal transformation (like a beam-splitter),
the unitary localization and the action of the decoherent map of
\eq{cmevo} commute, because $\sig_{\infty}\propto\id$ is obviously
preserved under orthogonal transformations (note that the bisymmetry
of the state is maintained through the channel, due to the symmetry
of the latter). In such cases, the decoherence of the bipartite
entanglement of the original three-mode state (with genuine
tripartite correlations) is exactly equivalent to that of the
corresponding initial two-mode state obtained by unitary
localization. This equivalence breaks down, even for GHZ/$W$ states
which can be localized through an (orthogonal) beam-splitter
transformation, for non homogeneous baths, {\em i.e.~} if the
thermal photon numbers $N_i$ related to different modes are
different [which is the case for different temperatures $T_i$ or for
different frequencies $\omega_i$, according to \eq{bose}] or if the
couplings $\gamma_i$ are different. In this instance let us remark
that unitary localization could provide a way to cope with
decoherence, limiting its hindering effect on entanglement. In fact,
let us suppose that a given amount of genuine tripartite
entanglement is stored in a symmetric (unitarily localizable)
three-mode state and is meant to be exploited, at some (later) time,
to implement tripartite protocols. During the period going from its
creation to its actual use such an entanglement decays under the
action of decoherence. Suppose the three modes involved in the
process do not decay with the same rate (different $\gamma_i$) or
under the same amount of thermal photons (different $N_i$), then the
obvious, optimal way to shield tripartite entanglement is
concentrating it, by unitary localization, in the two least
decoherent modes. The entanglement can then be redistributed among
the three modes by a reversal unitary operation, just before
employing the state. Of course, the concentration and distribution
of entanglement require a high degree of non local control on two of
the three-modes, which would not always be allowed in realistic
operating conditions.

The {\em bipartite}
entanglement of GHZ/$W$ states (under $(1+2)$-mode bipartitions)
decays slightly faster (in homogeneous baths with equal number of
photons) than that of an initial pure two-mode squeezed state (also
known as ``twin-beam'' state) with the same initial entanglement. In
this respect, multimode entanglement is more fragile than two-mode,
as the Hilbert space exposed to decoherence which contains it is
larger.
Notice that this claim does not refute the one of Ref.~\cite{paris05},
where SU(2,1) coherent states were found to be as robust as corresponding
two-mode states, but only for the same total number of thermal photons in the
multimode channels.

\section{Concluding remarks and outlook}

Gaussian states distinctively stand out in the infinite variety of
quantum states of continuous-variable systems, both for the analytic
description they allow in terms of covariance matrices and
symplectic operations, and for the high standards currently reached
in their experimental production, manipulation and implementation
for CV quantum information processing. Still, some recent results
demonstrate that basically the current state of the art in the
theoretical understanding and experimental control of CV
entanglement is strongly pushing towards the boundaries of the
``ideal'' realm of Gaussian states and Gaussian operations. For
instance, Gaussian entanglement cannot be distilled by Gaussian
operations alone \cite{nogo1,nogo2,nogo3}, and moreover Gaussian
states are ``extremal'', in the sense that they are the {\em least}
entangled among all states of CV systems with a given CM
\cite{wolfext}. On the other hand, however, some important pieces of
knowledge in the theory of entanglement of Gaussian states are still
lacking. The most important asymptotic measures of entanglement
endowed with a physical meaning, the entanglement cost and the
entanglement of distillation cannot be computed, and the
entanglement of formation is computable only in the special case of
two-mode, symmetric Gaussian states \cite{eofprl}. Moreover, when
moving to consider multipartite entanglement, many of the basic
questions are still unanswered, much like in the case of
multipartite entanglement in states of many qubits.

In this work we took a step ahead in the characterization of
multipartite entanglement in Gaussian states. We focused on the
prototypical structure of a CV system with more than two parties,
that is a three-mode system prepared in a Gaussian state. We
completed the elegant qualificative classification of separability
in three-mode Gaussian states provided in Ref. \cite{kraus} with an
exhaustive, quantitative characterization of the various forms of
quantum correlations that can arise among the three parties. We then
exploited some recent results on entanglement sharing in multimode
Gaussian states \cite{contangle} that prove that CV entanglement in
these states is indeed monogamous in the sense of the
Coffman-Kundu-Wootters monogamy inequality \cite{CKW}. We next
defined a measure of genuine tripartite entanglement, the residual
continuous-variable tangle, that turns out to be an entanglement
monotone under tripartite Gaussian LOCC \cite{contangle}.

We started our analysis by giving a complete characterization of
pure and mixed three-mode Gaussian states, and deriving the standard
forms of the covariance matrices that are similar to those known for
two-mode states \cite{duan00}. In particular, a generic pure
three-mode Gaussian states is completely specified, in standard
form, by three parameters, which are the purities (determinants of
the CMs) of the reduced states for each mode. We determined
analytically the general expression of the genuine tripartite
entanglement in pure three-mode Gaussian states, and studied its
properties in comparison with the bipartite entanglement across
different partitions. We investigated the sharing structure
underlying the distribution of quantum correlations among three
modes in arbitrary Gaussian states, much on the same lines as those
followed in the case of states of three qubits \cite{wstates}.

Remarkably, we found a completely unique feature, namely that that
there exists a special class of states, the pure, symmetric,
three-mode squeezed states, which simultaneously maximize the
genuine tripartite entanglement {\it and} the bipartite entanglement
in the reduced states of any pair of modes. This property, which has
no counterpart in finite-dimensional systems, can be understood as
the {\em promiscuous sharing} of CV entanglement. The states
exhibiting this peculiar sharing structure, named CV ``GHZ/$W$''
states for self-explaining reasons, are automatically preselected as
optimal carriers of quantum information over lossy channels, and we
proved that they indeed are. In fact, we concluded our work with a
detailed analysis of the effects of decoherence on three-mode
Gaussian states and the decay of tripartite entanglement. This study
yielded that the GHZ/$W$ states are the most robust three-party
entangled Gaussian states against decoherence.

We believe that the collection of results presented here, although
remarkable on its own, is however only the tip of an iceberg.
Three-mode Gaussian states, the perfect test-ground for the
understanding of some generic traits of multipartite entanglement in
CV systems, need to be analyzed in a deeper future perspective.
 This primarily includes the characterization of those
classes of tripartite entangled states with peculiar properties,
with a particular care towards their state engineering in quantum
optical settings. This analysis is currently under way
\cite{3mjpart1}. The (closely related) usefulness of such states for
existing and maybe novel protocols of CV quantum communication, able
to take advantage from the promiscuous sharing, is also being
investigated \cite{3mjpart2}.

From a broader theoretical standpoint, further research stemming from
the present work should probably be directed along two
main directions. The first one concerns proving a general
monogamy inequality in all multimode states of CV systems,
in analogy to what has been recently established
for arbitrary states of multiqubit systems \cite{Osborne}. Such
a proof would then lead to a multimode generalization of the residual
contangle. The second, long-term direction is the investigation
of the qualitative and quantitative aspects of entanglement in
generic non-Gaussian states of CV systems.
In this context, singling out exotic states with enhanced
promiscuous sharing of quantum correlations and
with a monogamy of entanglement stretched to its limits, appears as
an exciting perspective, and might open very promising perspectives for
the manipulation, transfer, and control of quantum information with
continuous variables.

\acknowledgments{Financial support from MIUR, INFN, and CNR is
acknowledged. AS acknowledges financial support from EPSRC, through the
QIP-IRC.}



\begin{thebibliography}{999}
\bibitem{Footnote} If $\sig$ is decomposed into pure non-Gaussian
states, it is not known at the present stage whether the CKW monogamy
inequality \eq{CKWine} is satisfied by each of them.
\bibitem{bornmarkov} Let us recall that such an approximation requires
to consider small couplings. In this way, the effect of $H_{int}$
can be truncated to the first order in the Dyson series. Moreover,
it neglects memory effects, in the sense that the future history of
the system depends only on its present state.
\end{thebibliography}
\end{document}